\documentclass[aps,pra,twocolumn,showpacs,superscriptaddress]{revtex4-1}

\usepackage{amsfonts,mathrsfs,amsmath,amsthm,amssymb}
\usepackage{bm,times}
\usepackage{graphicx,epsfig}
\usepackage[colorlinks=true,breaklinks=true,linkcolor=blue,citecolor=blue,urlcolor=blue]{hyperref}

\def \tr{ {\rm{Tr}}}
\def \non {\nonumber}

\newcommand{\be}{\begin{eqnarray}}
\newcommand{\ee}{\end{eqnarray}}
\newcommand{\dg}{\dagger}
\newcommand{\im}{\mathrm{i}}

\makeatletter
    
    \newcommand{\Rmnum}[1]{\expandafter\@slowromancap\romannumeral #1@}
\makeatother

\begin{document}
\title{Robustness of the concatenated quantum error-correction protocol against noise for channels affected by fluctuation}
\author{Long Huang}
\affiliation{College of Physical Science and Technology, Sichuan University, Chengdu 610064, China}
\author{Xiaohua Wu}
\email{wxhscu@scu.edu.cn}
\affiliation{College of Physical Science and Technology, Sichuan University, Chengdu 610064, China}
\author{Tao Zhou}
\email{taozhou@swjtu.edu.cn}
\affiliation{Department of Applied Physics, School of Physical Science and Technology, Southwest Jiaotong University, Chengdu 611756, China}

\begin{abstract}
In quantum error correction, the description of noise channel cannot be completely accurate, and fluctuation always appears in noise channel. It is found that when fluctuation of physical noise channel is considered, the average effective channel is dependent only on the average of physical noise channel, and the average of physical noise channel here plays the role of the independent error model in the previous works. Now, one may conclude that in the independent error model, the results in previous works are also valid for average channel where fluctuation exists. In some typical cases, our numerical simulations in the concatenated QEC protocol with 5-qubit code, 7-qubit Steane code and 9-qubit Shor confirm this conjecture. For 5-qubit code, the effective channels are approximate to depolarizing channel as the concatenated level increases. For Steane code, the effective channels are approximate to one Pauli channel as the concatenated level increases. For Shor code, the effective channels are approximate to one of Pauli-$X$ and Pauli-$Z$ channels in each level, and in next concatenated level, the effective channels are approximate to the other. Meanwhile, for these codes, the numerical results indicate that the degree of approximation increases with the concatenated level increases, and the fluctuation of noise channel decays exponentially as concatenated QEC performed. On the error-correction threshold, attenuation ratio of standard deviation of channel fidelity roughly has a stable value. On the other hand, standard deviations of off-diagonal elements of quantum process matrix (Pauli Form) decay more quickly than standard deviations of diagonal elements.
\end{abstract}


\date{\today}

\maketitle

\section{Introduction}

In quantum computation and communication, quantum error correction (QEC) was developed from classic schemes to preserve coherent states from noise and other unexpected interactions. Shor~\cite{Shor} introduced a strategy to store a bit of quantum information in an entanglement state of nine qubits, and Steane~\cite{Steane} proposed a protocol that uses seven qubits. The 5-qubit code was discovered by Bennett \textit{et al.}~\cite{Bennett} and independently by Laflamme \textit{et al.}~\cite{Laflamme}. Meanwhile, QEC conditions were proven independently by Bennett and co-authors~\cite{Bennett} and by Knill and Laflamme~\cite{KandL}. All the protocols with quantum error correction codes (QECCs) can be viewed as active error correction. Another way, the decoherence-free subspaces~\cite{Duan,Lidar,Zanardi} and noiseless subsystem~\cite{KandLV,Zanardi2,Kempe} are passive error-avoiding techniques.  Recently, it has been proven that both the active and passive QEC methods can be unified~\cite{Kribs,Poulin 05,Kribs2}.

The standard QEC procedure in Refs.~\cite{Steane,Bennett,Laflamme} is designed according to the principle of perfect correction for arbitrary single-qubit errors, where one postulates that single-qubit errors are the dominant terms in the noise process~\cite{Nielsen}. Recently, rather than correcting arbitrary single-qubit errors, the error recovery scheme was adapted to model for the noise to maximize the fidelity of the operation~\cite{Reimpell,Fletcher 07,Yamamoto,Fletcher 08}. When the uncertainty of the noise channel is considered, robust channel-adapted QEC protocols have also been developed~\cite{Kosut PRL,Kosut,Ball}. When the fidelity obtained from error correction is not high enough, the further increase in levels of concatenation is necessary. In the previous works~\cite{Poulin 06, Rahn, Kesting}, the concatenated code was discussed for the Pauli channel, where the depolarizing channel as the most important example is included, and quite recently, universal concatenated quantum codes have been well discussed by Chamberland \textit{et. al}~\cite{PRL117.010501,PRA95.022313}. In realization of QEC, it was revealed that non-degenerate stabilizer code with a complete set of fault-tolerant one-qubit Clifford gates always has a universal set of fault-tolerant gates~\cite{Theodore}. Moreover, error rates and resource overheads of some of the universal concatenated codes were analyzed~\cite{Ryuji}, and however, in this work, the ideal condition is considered, where errors do not exist in encoding and decoding operations. Before applying specific QEC operation for maximize the fidelity, we need to get the noise model by measuring Choi~matrix~\cite{Gilchrist,Emerson,Knill 08,Bendersky,Magesan 11,Magesan 12}. The standard quantum process tomography (QPT)~\cite{Nielsen,HengYan,Ariano03,Ariano04,Chuang,Poyatos,Leung,Childs,Shao-Ming} can be employed to determine Choi matrix of the effective channel, and the exact performance of concatenated QEC can be denoted by the effective Choi matrix.

One should notice that it is not effective enough to alter operations in each level of concatenated QEC for every measured noise model. Fortunately, concatenated QEC with 5-qubit code is general for correcting common noise models such as depolarizing, bit-flip, amplitude damping, even arbitrary kinds of noise model~\cite{Huang}. On the other hand, the noise in QEC is always approximately known~\cite{C.C2017}, and the estimations for noise process in experiments indicate that fluctuation exists in physical noise channel. Meanwhile, the interaction between the system and environment cannot be fully accurate due to its complexity~\cite{Ball} and the fluctuation in noise channel does exist based on experiment measurements~\cite{L. Steffen}. Therefore, the physical noise channel may be appropriately described by both the average and fluctuation of noise channels. So, it is necessary to consider the fluctuation of quantum noise channel in concatenated QEC protocol.

In this work, we study the performance of concatenated QEC protocol with 5-qubit code, 7-qubit Steane code and 9-qubit Shor code for noise channels with fluctuation, and the exact performance of QEC is represented by the quantum process matrix (QPM) in Pauli form, which is equivalent to Choi matrix. It is found that when fluctuation of physical noise channel is considered, the average effective channel is dependent only on the average of physical noise channel. In the following, we progress numerical simulations with specific noise models, and the results show that the average effective channel is indeed dependent only on the average of physical noise channel. Meanwhile, numerical results also indicate that the fluctuation of noise channel decays exponentially as concatenated QEC performed.

To be specific, the numerical simulations indicate that the standard deviations (SDs) of both the effective channel fidelity and diagonal elements of QPM decay exponentially as the concatenated level increases, and the SDs of off-diagonal elements of QPM (Pauli Form) decay even more quickly. For 5-qubit code, SDs of off-diagonal elements of QPM (Pauli form) decay at least $70$ times faster than SDs of diagonal elements, and the effective channels are approximate to depolarizing channel as the concatenated level increases. For Steane code, SDs of off-diagonal elements of QPM (Pauli form) decay at least $50$ times faster than SDs of diagonal elements, and the effective channels are approximate to a Pauli channel as the concatenated level increases. For Shor code, SDs of off-diagonal elements of QPM (Pauli Form) decay at least $30$ times faster than those of diagonal elements. The effective channels are approximate to either of Pauli-$X$ and Pauli-$Z$ channels in each level, and in next concatenated level, the effective channels are approximate to the other. Meanwhile, for these codes, the numerical results indicate that the degree of approximation increases as the concatenated level increases. On the error-correction threshold, attenuation ratio of SD of channel fidelity roughly has a stable value. On the other hand, for 5-qubit code and Steane code, different models of physical noise almost have no influence on the attenuation ratios of the SDs of channel fidelity, which are nearly only dependent on the average effective channel fidelity. While, for Shor code, different models of physical noise have obvious influence on the attenuation ratios of the SDs of channel fidelity.

The content of the present work is organized as follows. In Sec.~\ref{ii}, QEC protocol is introduced. In Sec.~\ref{iii}, we review QPT and characterize fluctuation in noise without QEC. In Sec.~\ref{iv}, we will show that the average effective channel after performing QEC is only dependent on the average of initial noise channels. In Sec.~\ref{v}, an exact noise model for numerical calculation is performed and the main results for 5-qubit code could be obtained. In Sec.~\ref{vi}, the main conclusions for 7-qubit Steane code and 9-qubit Shor code could be obtained. In Sec.~\ref{vii}, the analysis of numerical results has been given. We end this work with some remarks and discussion in Sec.~\ref{viii}.

\section{Quantum error correction protocol}
\label{ii}

\begin{figure}[tbph]
\centering
\includegraphics[width=0.47 \textwidth]{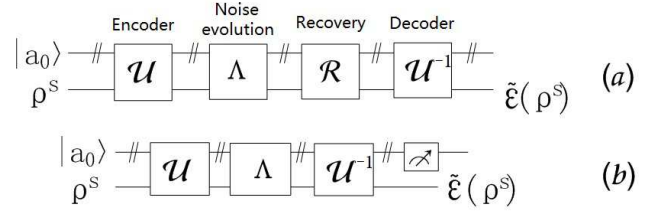}\\
\caption{(a) The way to construct the effective channel from the standard QEC protocol includes encoding, noise evolution, recovery, and decoding. (b) Our protocol where a unitary transformation $\mathcal{U}$ is sufficient to correct the errors of the principle system. The errors of the assistant qubits are left uncorrected. }\label{figure1}
\end{figure}

In recent works, QEC is developed to preserve coherent states from noise and other unexpected interactions. As depicted in Fig.~\ref{figure1}~(a), the standard way to obtain the effective noise channel contains the following steps: (i) A unitary transformation $U$ for encoding process $\mathcal{U}$; (ii) The noise evolution denoted by $\Lambda$; (iii) The recovery operation described by a process $\mathcal{R}$; (iv) The decoding process $\mathcal{U}^{-1}$ realized by $U^{\dag}$.

In this paper, the protocol shown in Fig.\ref{figure1} (b) is used,
we start the QEC protocol with 5-qubit code in Ref.~\cite{Bennett},
\be
|0_\mathcal{L}\rangle&=&\frac{1}{4}[|00000\rangle+|10010\rangle+|01001\rangle+|10100\rangle\non\\
&&+|01010\rangle-|11011\rangle-|00110\rangle-|11000\rangle\non\\
&&-|11101\rangle-|00011\rangle-|11110\rangle-|01111\rangle\non\\
&&-|10001\rangle-|01100\rangle-|10111\rangle+|00101\rangle],\non
\ee
and
\be
|1_\mathcal{L}\rangle&=&\frac{1}{4}[|11111\rangle+|01101\rangle+|10110\rangle+|01011\rangle\non\\
&&+|10101\rangle-|00100\rangle-|11001\rangle-|00111\rangle\non\\
&&-|00010\rangle-|11100\rangle-|00001\rangle-|10000\rangle\non\\
&&-|01110\rangle-|10011\rangle-|01000\rangle+|11010\rangle].\non
\ee
In this scheme, the special parts are the unitary process $\mathcal{U}$ and its associated process $\mathcal{U}^{-1}$, which work not only as encoding and decoding, but also as error correction. [The process $\mathcal{U}^{-1}$ here is just the $U_2$ used in Eq. (87) of the original work in Ref.~\cite{Bennett}].The specific unitary process $\mathcal{U}$ is designed as
\be
\label{d1}
U|a_{m}\rangle\otimes|0\rangle=E_{m}|0_\mathcal{L}\rangle,\non\\
U|a_{m}\rangle\otimes|1\rangle=E_{m}|1_\mathcal{L}\rangle,\non\\
U^{\dagger}E_{m}|0_\mathcal{L}\rangle=|a_{m}\rangle\otimes|0\rangle,\non\\
U^{\dagger}E_{m}|1_\mathcal{L}\rangle=|a_{m}\rangle\otimes|1\rangle,
\ee
where $m=0,1,...,15$, $E_0$ is the identity operator, and for $m\neq0$, $E_{m}$ is one of the Pauli operators $\sigma^{i}_{j}(i=1,...5, j=1,2,3)$.

According to the analysis in Ref.~\cite{Bennett}, the recovery process $\mathcal{R}$ in Fig.~\ref{figure1}~(a) is not necessary and can be moved away, since the $\mathcal{U}^{-1}$ defined in Eq.~(\ref{d1}) is sufficient for correcting the errors of the principle system. One can observe that the following two processes are equivalent
\be
\mathcal{R}\circ\mathcal{U}^{-1}\equiv\mathcal{U}^{-1}\circ\tilde{\mathcal{R}},\non
\ee
with the new process,
\be
\tilde{\mathcal{R}}=\mathcal{U}\circ\mathcal{R}\circ\mathcal{U}^{-1}.\non
\ee
Furthermore, it can be expressed in a more explicit way as ${\mathcal{\tilde{R}}(\rho^{\mathcal{SA}}})=\sum^{15}_{m=0}\tilde{R}_m\rho^{SA}\tilde{R}^{\dag}_{m}$, with the Kraus operators $\tilde{R}_m=UR_m{U}^\dg$. By some simple algebra, one can get $\tilde{R}_m=|a_{0}\rangle\langle a_{m}|\otimes I$, and the state of the principle system remains unchanged. Therefore, when the protocol is applied in quantum information storage and transmission, the recovery of auxiliary qubits $\tilde{R}_m$ can be abandoned. In this work, the encoding process $\mathcal{U}$ and the decoding (recovery) process $\mathcal{U}^{-1}$ are fixed in every level of the concatenated QEC.

\section{Quantum process matrix for noise with fluctuation}
\label{iii}

In this work, QPT can be an accurate tool to characterize the performance of QEC protocol in Sec.~\ref{ii}. Before giving a brief review of the general theory about QPT, one can first introduce a convenient tool where a bounded operator on a Hilbert space can be associated with a vector in an extended Hilbert space. Let $A$ be a bounded operator in a $d$-dimensional Hilbert space $H_{d}$, with $A_{ij}=\langle i|A|j\rangle$ the matrix elements, and an isomorphism between $A$ and a vector $|A\rangle\rangle$ in $H^{\otimes2}_{d}$ is defined as
\be
\label{a00}
|A\rangle\rangle=\sqrt{d}A\otimes I_{d}|S_{+}\rangle=\sum^{d}_{i,j=1}A_{ij}|ij\rangle,
\ee
where $|S_{+}\rangle=\frac{1}{\sqrt{d}}\sum^{d}_{k=1}|kk\rangle$ is a maximally entangled state in $H^{\otimes2}_{d}$, and $|ij\rangle=|i\rangle\otimes|j\rangle$. This isomorphism offers a one-to-one map between an operator and its vector form. Suppose
that $A$, $B$ and $\rho$ are three arbitrary bounded operators in $H_{d}$, and then
\be
\label{a01}
\tr[A^{\dagger}B]=\langle\langle A|B\rangle\rangle, \ |A\rho B\rangle\rangle=A\otimes B^{T}|\rho\rangle\rangle,
\ee
with $B^{T}$ the transpose of $B$.

For a noise channel with fluctuation, a set of parameters $\bm\omega=\{\omega_1,\omega_2,...\omega_n\}$ can be introduced to represent the noise, and the quantum channel can be described by a set of Kraus operators $\{E_{m}(\bm\omega)\}$,
\be
\label{a1}
\varepsilon{(\bm\omega)}[\rho]=\sum_{m}E_{m}(\bm\omega)\rho E^{\dagger}_{m}(\bm\omega).
\ee
For instance, a unitary channel on a one-qubit system can be expressed as
\be
\label{a2}
\varepsilon{(\bm\omega)}[\rho]&=&U(\bm\omega)\rho U^{\dagger}(\bm\omega),
\ee
where $U(\bm\omega)=\cos\theta I+\im\sin\theta \bm{\sigma}\cdot \hat{n}(\gamma,\phi)$, $\bm\omega=\{\theta,\gamma,\phi\}$, $\theta,\gamma\in[0,2\pi)$, $\phi\in[0,\pi]$, and $\hat{n}=(\sin\phi\cos\gamma,\sin\phi\sin\gamma,\cos\phi)$.

Now, for a quantum channel $\varepsilon{(\bm\omega)}$, Choi-Jamiolkowski isomorphism is a useful connection between a quantum channel and a bipartite state
\be
\label{a3}
\hat{\chi}(\bm\omega)&=&d\cdot\varepsilon{(\bm\omega)}\otimes I_{d}(|S_{+}\rangle\langle S_{+}|)\non\\
&=&\sum_{m}|E_{m}(\bm\omega)\rangle\rangle\langle\langle E_{m}(\bm\omega)|,
\ee
where $\hat{\chi}(\bm\omega)$ is the so-called Choi matrix, and it can be measured in experiment with assistant channel. Therefore, according to Eq.~(\ref{a00}) and Eq.~(\ref{a01}), Eq.~(\ref{a1}) can be rewritten as
\be
|\varepsilon{(\bm\omega)}[\rho]\rangle\rangle&=&\sum_{m}|E_{m}(\bm\omega)\rho E^{\dagger}_{m}(\bm\omega)\rangle\rangle\non\\
&=&\sum_{m}E_{m}(\bm\omega)\otimes E^{\ast}_{m}(\bm\omega)|\rho\rangle\rangle\non\\
&=&\hat{\lambda}(\bm\omega)|\rho\rangle\rangle\non,
\ee
with
\be
\label{a4}
\hat{\lambda}(\bm\omega)\equiv\sum_{m}E_{m}(\bm\omega)\otimes E^{\ast}_{m}(\bm\omega)
\ee
the quantum process matrix (QPM). So, in the general theory of QPT, a quantum channel can be equivalently represented by the corresponding QPM. From Ref.~\cite{X.-H. Wu}, the relationship between Choi matrix $\hat{\chi}_{ab;cd}(\bm\omega)=\langle ab|\hat{\chi}(\bm\omega)|cd\rangle$ and QPM $\hat{\lambda}_{ab;cd}(\bm\omega)=\langle a|\varepsilon(\bm\omega)[|c\rangle\langle d|]|b\rangle (a,b,c,d=0,1)$ can be expressed as
\be
\label{a5}
\hat{\chi}_{ab;cd}(\bm\omega)=\hat{\lambda}_{ac;bd}(\bm\omega).
\ee

For one-qubit case, a quantum state can be repressed in Bloch representation
\be
\rho=\frac{1}{2}(\hat{\sigma_0}+\bm r\cdot\bm{\hat{\sigma}}),\non
\ee
where $\hat{\sigma}_0$ is the identity operator, $\hat{\bm{\sigma}}=(\hat{\sigma}_1,\hat{\sigma}_2,\hat{\sigma}_3)$ are Pauli operators, and $\bm r$ is Bloch vector. Meanwhile, the quantum process in Eq.~(\ref{a1}) for qubit system can now be represented in Bloch representation  as
\be
\bm{r}'(\bm\omega)= \hat{M}(\bm\omega)\bm{r}+\bm{c}(\bm\omega),\non
\ee
with $\bm{r}'(\bm\omega)$ a new Bloch vector, $\hat{M}(\bm\omega)$ a $3\times3$ real matrix, and $\bm{c}(\bm\omega)$ a constant vector. Now, QPM in Pauli basis can be introduced to character the quantum noise process more clearly,
\be
\label{a7}
\eta_{\mu\nu}(\bm\omega)&=&\tr[\sigma^{\dagger}_{\mu}\varepsilon{(\bm\omega)}[\sigma_{\nu}]]=\langle\langle\sigma_{\mu}|\varepsilon{(\bm\omega)}[\sigma_{\nu}]\rangle\rangle \non\\
&=&\langle\langle\sigma_{\mu}|\hat{\lambda}(\bm\omega)|\sigma_{\nu}\rangle\rangle,
\ee
where $\sigma_{\mu},\sigma_{\nu}=\frac{\sqrt{2}}{2}\hat{\sigma}_0,\frac{\sqrt{2}}{2}\hat{\sigma}_1,\frac{\sqrt{2}}{2}\hat{\sigma}_2,\frac{\sqrt{2}}{2}\hat{\sigma}_3$ for $\mu,\nu=0,1,2,3$. With the fact that the quantum channel $\varepsilon(\bm\omega)$ is always trace-preserving, it is easy to obtain
\be
\label{a6}
\eta(\bm\omega)=\left(\begin{array}{cc}
1 & 0\\
\bm c(\bm\omega) & \hat{M}(\bm\omega) \end{array}
\right) .
\ee
Since the QPM in Pauli basis is more compact than that in Eq.~(\ref{a4}), we mainly discuss the QPM in Eq.~(\ref{a6}) in the following.

In experiments, the measurements of QPM are not fully accurate, and the fluctuation always exists in the measurements~\cite{L. Steffen}. Therefore, in QEC, one should take fluctuation of noise into account. In statistics, the average of the element $\eta_{\mu\nu}(\bm\omega)$ of QPM $\eta(\bm\omega)$ can be obtained
\be
\label{a8}
\eta^{\mathrm{avg}}_{\mu\nu}&=&\int p(\bm\omega)d\bm\omega \eta_{\mu\nu}(\bm\omega) \non\\
&=&\int p(\bm\omega)d\bm\omega\tr\big(\sigma^{\dagger}_{\mu}\varepsilon{(\bm\omega)}[\sigma_{\nu}]\big) \non\\
&=&\tr\big(\sigma^{\dagger}_{\mu}\varepsilon^{\mathrm{avg}}[\sigma_{\nu}]\big),
\ee
where $p(\bm\omega)=p(\omega_{1})p(\omega_{2})...p(\omega_{n})$, $d\bm\omega=d\omega_{1}d\omega_{2}...d\omega_{n}$, and the average QPM $\eta^{\mathrm{avg}}$ corresponds to the average noise channel $\varepsilon^{\mathrm{avg}}=\int p(\bm\omega)d\bm\omega\varepsilon{(\bm\omega)}$. For the example in Eq.~(\ref{a2}), $p(\bm\omega)=\sin\phi/8\pi^{2}$, $d\bm\omega=d\theta d\gamma d\phi$, $\theta,\gamma\in[0,2\pi]$, $\phi\in[0,\pi]$. The average QPM can be obtained
\begin{equation}
\label{au}
\eta^{\mathrm{avg}}=\left(\begin{array}{cccc}
1 & 0& 0& 0\\
0 & \frac{1}{3}& 0& 0\\
0 & 0& \frac{1}{3}& 0\\
0 & 0& 0& \frac{1}{3}\\\end{array}
\right) ,
\end{equation}
and the average channel $u^{\mathrm{avg}}$ can be represented by a set of Kraus operators: $\{\sqrt{\frac{1}{2}}\hat{\sigma}_{0}, \sqrt{\frac{1}{6}}\hat{\sigma}_{1}, \sqrt{\frac{1}{6}}\hat{\sigma}_{2}, \sqrt{\frac{1}{6}}\hat{\sigma}_{3}\}$.

In order to characterize the fluctuation in noise channels, the SD of element $\eta_{\mu\nu}(\bm\omega)$ is defined
\be
\label{a9}
\delta\eta_{\mu\nu}=\sqrt{\int p(\bm\omega)d\bm\omega[\eta_{\mu\nu}(\bm\omega)-\eta^{\mathrm{avg}}_{\mu\nu}]^{2}}.
\ee

Meanwhile, as an important characterization of noise channel $\varepsilon{(\bm\omega)}$, channel fidelity $F(\bm\omega)$~\cite{Schumacher} can be obtained from QPM $\eta(\bm\omega)$,
\be
\label{a10}
F(\bm\omega)=\frac{1}{4}\tr[\eta(\bm\omega)].
\ee
In statistics, the average channel fidelity $F^{\mathrm{avg}}$ can be obtained
\be
\label{a11}
F^{\mathrm{avg}}&=&\int p(\bm\omega)d\bm\omega F(\bm\omega) \non\\
&=&\int p(\bm\omega)d\bm\omega\frac{1}{4}\tr[\eta(\bm\omega)] \non\\
&=&\frac{1}{4}\tr(\eta^{\mathrm{avg}}),
\ee
and the fluctuation of channel fidelity $F(\bm\omega)$ is characterized by the SD of channel fidelity
\be
\label{a12}
\delta F=\sqrt{\int p(\bm\omega)d\bm\omega[F(\bm\omega)-F^{\mathrm{avg}}]^{2}}.
\ee

\begin{figure}[tbph]
\centering
\includegraphics[width=0.4 \textwidth]{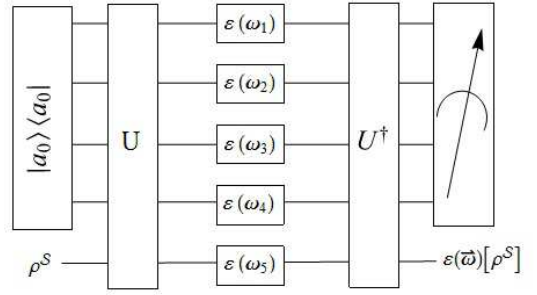}\\
\caption{ The effective quantum process with QEC. }\label{figure2}
\end{figure}

In the following, to show the changes of SD of channel fidelity and SDs of elements of QPM, we define the SDs for the effective channel after performing QEC.

\section{The average effective channel after performing QEC}
\label{iv}

In this section, we will show that the average effective channel is only dependent on the average of initial noise channels. The effective quantum process for QEC is shown in Fig.~\ref{figure2}. The assistant qubits system is denoted by $\mathcal{A}$, and $|a_0\rangle$ is the ground state of 4 assistant qubits. The transformation $\mathcal{U}$ is the encoding operation, and it is shown in Sec.~\ref{ii} that $\mathcal{U}^{-1}$ is both the decoding and error correcting operation. $\bm\omega_i$ is a set of parameters of the physical channel for each qubit, and $\overrightarrow{\bm\omega}=\{\bm\omega_{1},\bm\omega_{2},\bm\omega_{3},\bm\omega_{4},\bm\omega_{5}\}$ is a set of independent parameters of noise channels for five qubit. $\Lambda(\overrightarrow{\bm\omega})=\otimes_{i=1}^5\varepsilon(\bm\omega_i)$ represents the noise process on five physical qubits. Now, the effective channel after QEC is denoted by $\varepsilon^{(1)}{(\overrightarrow{\bm\omega})}$.

As shown in Fig.~\ref{figure2}, after QEC is performed, the element $\eta^{(1)}_{\mu\nu}(\overrightarrow{\bm\omega})$ in the QPM ${\eta}^{(1)}(\overrightarrow{\bm\omega})$ of effective channel $\varepsilon^{(1)}{(\overrightarrow{\bm\omega})}$ is defined as
\be
\label{a13}
\eta_{\mu\nu}^{(1)}(\overrightarrow{\bm\omega})&=&\tr[\sigma^{\dagger}_{\mu}\tr_{\mathcal{A}}[\mathcal{U}^{-1}\circ\Lambda(\overrightarrow{\bm\omega})\circ\mathcal{U}(|a_0\rangle \langle a_0|\otimes\sigma_{\nu})]] \non\\
&=&\tr[\sigma^{\dagger}_{\mu}\tr_{\mathcal{A}}[\mathcal{U}^{-1}\circ\bigotimes_{i=1}^5\varepsilon{(\bm\omega_i)}\circ\mathcal{U}(|a_0\rangle \langle a_0|\otimes\sigma_{\nu})]].\non\\
\ee
In statistics, the average element $\eta^{\mathrm{avg}(1)}_{\mu\nu}$ of effective QPM ${\eta}^{(1)}(\overrightarrow{\bm\omega})$ is
\begin{widetext}
\be
\label{a14}
\eta^{\mathrm{avg}(1)}_{\mu\nu}&=&\int p(\overrightarrow{\bm\omega})d\overrightarrow{\bm\omega}\eta^{(1)}_{\mu\nu}(\overrightarrow{\bm\omega})=\int p(\overrightarrow{\bm\omega})d\overrightarrow{\bm\omega}\tr[\sigma^{\dagger}_{\mu}\tr_{\mathcal{A}}[\mathcal{U}^{-1}\circ\Lambda(\overrightarrow{\bm\omega})\circ\mathcal{U}(|a_0\rangle \langle a_0|\otimes\sigma_{\nu})]] \non\\
&=&\int \prod_{i=1}^5[p(\bm\omega_i)d\bm\omega_i]\tr\big[\sigma^{\dagger}_{\mu}\tr_{\mathcal{A}}[\mathcal{U}^{-1}\circ\bigotimes_{j=1}^5\varepsilon{(\bm\omega_j)}\circ \mathcal{U}(|a_0\rangle \langle a_0|\otimes\sigma_{\nu})]\big] \non\\
&=&\tr\big[\sigma^{\dagger}_{\mu}\tr_{\mathcal{A}}[\mathcal{U}^{-1}\circ {(\varepsilon^{\mathrm{avg}}})^{\otimes5}\circ\mathcal{U}(|a_0\rangle \langle a_0|\otimes\sigma_{\nu})]\big],
\ee
\end{widetext}
where $p(\overrightarrow{\bm\omega})=\Pi_{i=1}^5p(\omega_i)$ and $d\overrightarrow{\bm\omega}=\Pi_{i=1}^5d\bm\omega_i$. Similarly, for any quantum code, the average effective channel is dependent only on average of the initial noise channel.

Now, the SD of element $\eta^{(1)}_{\mu\nu}(\overrightarrow{\bm\omega})$ is defined as
\be
\label{a15}
\delta\eta^{(1)}_{\mu\nu}=\sqrt{\int p(\overrightarrow{\bm\omega})d\overrightarrow{\bm\omega}[\eta^{(1)}_{\mu\nu}(\overrightarrow{\bm\omega})-\eta^{\mathrm{avg}(1)}_{\mu\nu}]^{2}}.
\ee

Meanwhile, channel fidelity $F^{(1)}(\overrightarrow{\bm\omega})$ of effective noise channel $\varepsilon^{(1)}{(\overrightarrow{\bm\omega})}$ can be obtained from ${\eta}^{(1)}(\overrightarrow{\bm\omega})$,
\be
\label{a16}
F^{(1)}(\overrightarrow{\bm\omega})=\frac{1}{4}\tr[{\eta}^{(1)}(\overrightarrow{\bm\omega})].
\ee
In statistics, the average effective channel fidelity $F^{\mathrm{avg}(1)}$ is
\be
\label{a17}
F^{\mathrm{avg}(1)}&=&\int p(\overrightarrow{\bm\omega})d\overrightarrow{\bm\omega}F^{(1)}(\overrightarrow{\bm\omega}) \non\\
&=&\int p(\overrightarrow{\bm\omega})d\overrightarrow{\bm\omega}\frac{1}{4}\tr[{\eta}^{(1)}(\overrightarrow{\bm\omega})] \non\\
&=&\frac{1}{4}\tr[{\eta}^{\mathrm{avg}(1)}] ,
\ee
and the SD of effective channel fidelity $F^{(1)}(\overrightarrow{\bm\omega})$ is defined
\be
\label{a18}
\delta F^{(1)}=\sqrt{\int p(\overrightarrow{\bm\omega})d\overrightarrow{\bm\omega}[F^{(1)}(\overrightarrow{\bm\omega})-F^{\mathrm{avg}(1)}]^{2}}.
\ee

The strict results for Eq.~(\ref{a15}) and Eq.~(\ref{a18}) are difficult to obtain. However, in Eq.~(\ref{a14}), we have proved the average elements of effective QPM $\eta^{\mathrm{avg}(1)}_{\mu\nu}$ are just determined by the average of 5 independent initial noise channels, and the strict results for Eq.~(\ref{a14}) and Eq.~(\ref{a17}) can be obtained. Obviously, in concatenated QEC protocol, the average elements of effective QPM and the average effective channel fidelity are just determined by the average of $n=5^{l}$ independent initial noise channels, where $l$ is the concatenated level. Due to the assumption of independent noise model, the average of independent initial noise channels in each qubit have the same expression, and $n=5^{l}$ sets of parameters $\{\bm\omega_{1},\bm\omega_{2},...\bm\omega_{5^{l}}\}$ also belong to the same distribution. When the $n=5^{l}$ initial noise channels are set as ${\varepsilon^{\mathrm{avg}}}^{\otimes5^{l}}$,
the average effective channel $\varepsilon^{\mathrm{avg}(l)}$ (with channel fidelity $F^{\mathrm{avg}(l)}$ and QPM $\eta^{\mathrm{avg}(l)}$) in the $l$-th level of concatenation can be obtained strictly.

The initial noise channels are set as $\varepsilon^{\mathrm{avg}}$ in~\cite{Huang}, and it is indicated that the effective channels can be transformed to depolarizing channel quickly (after a two or three-level concatenated QEC), which behaves like twirling procedure~\cite{Horodecki}. According to this, one could conjecture that the fluctuation of initial noise channels will decay as the concatenated QEC performed, and in the following, numerical simulations for typical noise model are performed to affirm this conjecture.

\section{Quantum noise model and main results of 5-qubit code}
\label{v}
In the section, before performing numerical simulations with 5-qubit code, it should be noted that the noise in QEC is always approximately known, and the estimations for noise process in experiments indicate that fluctuation exists in physical noise channel. Therefore, it is more appropriate to describe physical noise channel by both the average and fluctuation of noise channels. To investigate the fluctuation of noise channels in concatenated 5-qubit QEC protocol, a perturbed noise model can be introduced,
\begin{equation}
\label{b1}
\varepsilon{(\bm\omega)}[\rho]=(1-k)\mathcal{N}[\rho]+ku(\bm\omega)[\rho],
\end{equation}
where $\mathcal{N}$ is a one-parameter noise model with channel fidelity $f\in[0,1]$, $k\in[0,1]$ is a small constant, and the unitary channel $u{(\bm\omega)}$ is expressed in Eq.~(\ref{a2}). This model is a simplified version of the noise model used to discuss the robustness of hard decoding algorithm in Ref.~\cite{C.C2017}, and the fluctuation is from the randomly distributed parameters $\bm\omega$ in $u{(\bm\omega)}$. On the other hand, based on estimations for fidelity in Ref.~\cite{L. Steffen}, the fluctuation of physical noise channel is small. Here, the fluctuation of physical noise channel is mainly determined by the constant $k$. Therefore, in our numerical simulations, $k$ is set bellow $10\%$ to ensure the fluctuation is small enough.

Now, from the definition in Eq.~(\ref{a7}), the QPM of $\varepsilon{(\bm\omega)}$ can be obtained,
\begin{equation}
\label{b0}
{\eta}(\bm\omega)=(1-k)\eta(\mathcal{N})+k\eta[u(\bm\omega)].
\end{equation}
When averaged on the parameters $\theta$,$\gamma$,$\phi$, as shown in Eq.~(\ref{au}), the average QPM of $u{(\bm\omega)}$ can be obtained, and the average QPM of $\varepsilon{(\bm\omega)}$ can be expressed as
\begin{equation}
\label{b2}
{\eta}^{\mathrm{avg}(0)}=(1-k)\eta(\mathcal{N})+k\left(\begin{array}{cccc}
1 & 0& 0& 0\\
0 & \frac{1}{3}& 0& 0\\
0 & 0& \frac{1}{3}& 0\\
0 & 0& 0& \frac{1}{3}\\\end{array}
\right).
\end{equation}
Meanwhile, the average of the perturbed noise channels in Eq. (\ref{b1}) can be obtained,
\begin{equation}
\label{b3}
\varepsilon^{\mathrm{avg}(0)}=(1-k)\mathcal{N}+ku^{\mathrm{avg}}.
\end{equation}

Now, as shown in Fig.~\ref{figure2}, one can perform QEC in each concatenated level $l$, and the average effective channel $\varepsilon^{\mathrm{avg}(l)}$ (with channel fidelity $F^{\mathrm{avg}(l)}$ and QPM $\eta^{\mathrm{avg}(l)}$) can be obtained, which can be used as the contrast in our numerical calculation.

More specifically, in our numerical calculation, $N_{0}$ is the number of sample noise channels in Eq.~(\ref{b1}), which is generated independently and randomly according to Eq.~(\ref{a2}). By performing concatenated 5-qubit QEC protocol, in the $l$-th level of concatenation, there are $N_{l}=N_{0}/5^{l}$ effective noise channels, where the average element $\bar{\eta}^{(l)}_{\mu\nu}$ and SD of element $\delta\eta^{(l)}_{\mu\nu}$ can be calculated by
\be
\label{b4}
\bar{\eta}^{(l)}_{\mu\nu}&=&\frac{1}{N_{l}}\sum^{N_{l}}_{i=1}[{\eta}^{(l)}_{i}]_{\mu\nu},\\
\label{b5}
\delta\eta^{(l)}_{\mu\nu}&=&\sqrt{\frac{1}{N_{l}}\sum^{N_{l}}_{i=1}[[{\eta}^{(l)}_{i}]_{\mu\nu}-\eta^{\mathrm{avg}(l)}_{\mu\nu}]^{2}}.
\ee
Here, ${\eta}^{(l)}_{i}$ represents the $i$-th ($i=1,2,...,N_l$) QPM of effective noise channels in the $l$-th level of concatenation.

Meanwhile, the average effective channel fidelity $\bar{F}^{(l)}$ and SD of channel fidelity $\delta F^{(l)}$ can be calculated according to
\be
\label{b6}
\bar{F}^{(l)}&=&\frac{1}{N_{l}}\sum^{N_{l}}_{i=1}F^{(l)}_{i}=\frac{1}{4N_{l}}\sum^{N_{l}}_{i=1}\tr[{\eta}^{(l)}_{i}] ,\\
\label{b7}
\delta F^{(l)}&=&\sqrt{\frac{1}{N_{l}}\sum^{N_{l}}_{i=1}[F^{(l)}_{i}-F^{\mathrm{avg}(l)}]^{2}}.
\ee
Here, we should note the average element of QPM used in Eq.~(\ref{b5}) and the average channel fidelity used in Eq.~(\ref{b7}) are obtained from concatenated QEC protocol for average initial noise channel in Eq.~(\ref{b3}), and both of them are strict, therefore the sample freedom should be $N_{l}$ rather than $N_{l}-1$.

To show changes of the fluctuation of sample noise channels under concatenated QEC protocol, we define attenuation ratio of SDs,
\be
\label{b8}
R^{(l)}_{F}=\frac{\delta F^{(l-1)}}{\delta F^{(l)}}, R^{(l)}_{\mu\nu}=\frac{\delta\eta^{(l-1)}_{\mu\nu}}{\delta\eta^{(l)}_{\mu\nu}}.
\ee

For the noise channel in Eq.~(\ref{b1}), there are $4$ variable factors: the noise model of $\mathcal{N}$,  channel fidelity $f$, the proportionality constant $k$, and an arbitrary unitary channel $u(\bm\omega)$. In the following, by generating unitary channels independently and randomly (the three parameters of $u(\bm\omega)$ are picked uniformly at random, and $\theta,\gamma\in[0,2\pi]$, $\phi\in[0,\pi]$) in concatenated 5-qubit QEC, the influences of the first three factors and the level $l$ in concatenated QEC on the fluctuation of noise channels are studied.

(i) For studying the impact of the noise model of $\mathcal{N}$ on the fluctuation, we set $f=0.98, k=0.02$, and the noise model of $\mathcal{N}$ is chosen as depolarizing noise, amplitude damping noise, and other 20 arbitrarily generated numerical noises, respectively. In each case $N_{0}=50000$ unitary channels are generated independently and randomly, and then one-level QEC are performed and numerical results are obtained.

\begin{figure}[tbph]
\centering
\includegraphics[width=0.46 \textwidth]{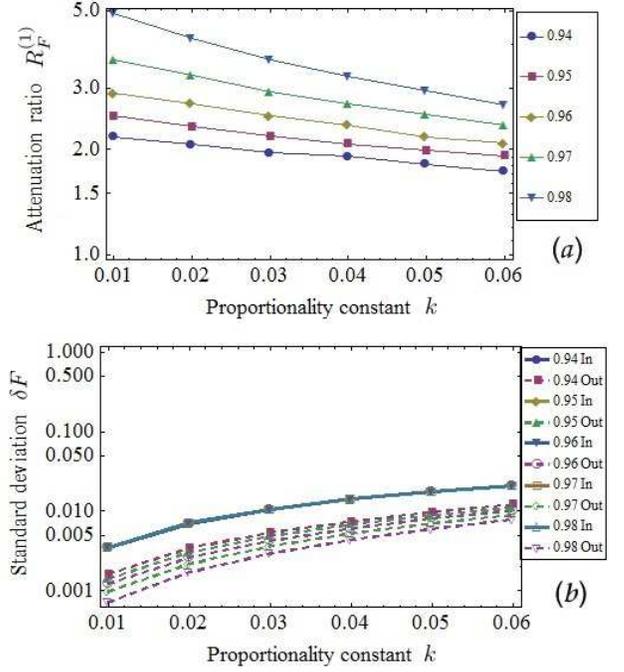}\\
\caption{(Color online) One-level QEC is performed, when $\mathcal{N}$ is set as amplitude damping noise: (a) Numerical results of attenuation ratio of SD of channel fidelity in Eq.~(\ref{b8}) ($l=1$) are depicted. The channel fidelity $f$ is set as $0.94, 0.95, 0.96, 0.97$, and $0.98$, respectively. The proportionality constant $k$ increases from $0.01$ to $0.06$ with a step $0.01$.
 (b) Numerical results of SD of channel fidelity in Eq.~(\ref{b7}) ($l=0,1$) are depicted. The channel fidelity $f$ is set as $0.94, 0.95, 0.96, 0.97$, and $0.98$, respectively. The proportionality constant $k$ increases from $0.01$ to $0.06$ with a step $0.01$. ``In'' means the case without performing QEC and ``Out'' means the case after performing one-level QEC.}\label{figure4}
\end{figure}

Numerical results indicate that the noise model of $\mathcal{N}$ has nearly no influence on both the SDs in Eqs.~(\ref{b5},\ref{b7}) and the average channel fidelity in Eq.~(\ref{b6}). In all cases, the SD of channel fidelity decays with attenuation ratio $R^{(1)}_{F}\approx4$, and the SDs of diagonal elements ($\delta\eta^{(1)}_{11},\delta\eta^{(1)}_{22},\delta\eta^{(1)}_{33}$) in QPM decay with attenuation ratio $R^{(1)}_{\mu\nu}\approx5 (\mu=\nu=1,2,3)$. The SDs of off-diagonal elements ($\delta\eta^{(1)}_{21}$, $\delta\eta^{(1)}_{31}$, $\delta\eta^{(1)}_{12}$, $\delta\eta^{(1)}_{32}$, $\delta\eta^{(1)}_{13}$, $\delta\eta^{(1)}_{23}$) decay more quickly (at least $200$ times faster) than the SDs of diagonal elements, and the SDs of first column elements ($\delta\eta^{(1)}_{10},\delta\eta^{(1)}_{20},\delta\eta^{(1)}_{30}$) transform from $0$ to a small amount (about $10^{-6}$) except the case where $\mathcal{N}$ is chosen as depolarizing noise.

(ii) On the other hand, we study the impact of $f$ and $k$ on the fluctuation. First, the noise model of $\mathcal{N}$ is chosen as an amplitude damping noise, and then we set $f=0.94,0.95,0.96,0.97,0.98$ respectively, and for each $f$, $k$ increases from $0.01$ to $0.06$ with a step $0.01$. In each case, $N_{0}=50000$ unitary channels are generated independently and randomly, and then one-level QEC are performed and numerical results can be obtained.

\begin{figure}[tbph]
\centering
\includegraphics[width=0.46 \textwidth]{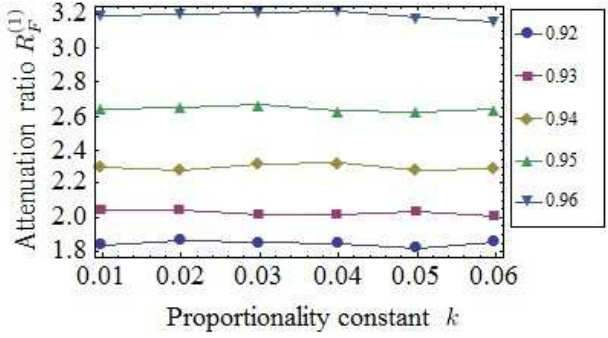}\\
\caption{(Color online) $\mathcal{N}$ is an amplitude damping channel, and $F^{\mathrm{avg}(0)}$ is set as $0.92,0.93,0.94,0.95,$ and $0.96$, respectively. For each $F^{\mathrm{avg}(0)}$, proportionality constant $k$ increases from $0.01$ to $0.06$ with a step $0.01$. After one-level QEC is performed, numerical results of attenuation ratio of SD of channel fidelity in Eq.~(\ref{b8}) ($l=1$) are depicted. }\label{figure5}
\end{figure}

Numerical results for the SD of channel fidelity are shown in Fig.~\ref{figure4}, and all results indicate that for initial noise channels, the SD of channel fidelity and the SDs of elements of QPM are almost only dependent on $k$. For fixed $f$, attenuation ratios of SDs after QEC are decreasing as $k$ increases. For fixed $k$, attenuation ratios of SDs after QEC are increasing as $f$ increases. In all cases, the SDs of off-diagonal elements of QPM decay more quickly (at least $70$ times faster) than the SDs of diagonal elements (the minimum attenuation ratio of the SDs of diagonal elements is about 2.15), and the SDs of first column elements transform from $0$ to a small amount (from $10^{-4}$ to $10^{-6}$).

\begin{figure}[tbph]
\centering
\includegraphics[width=0.47 \textwidth]{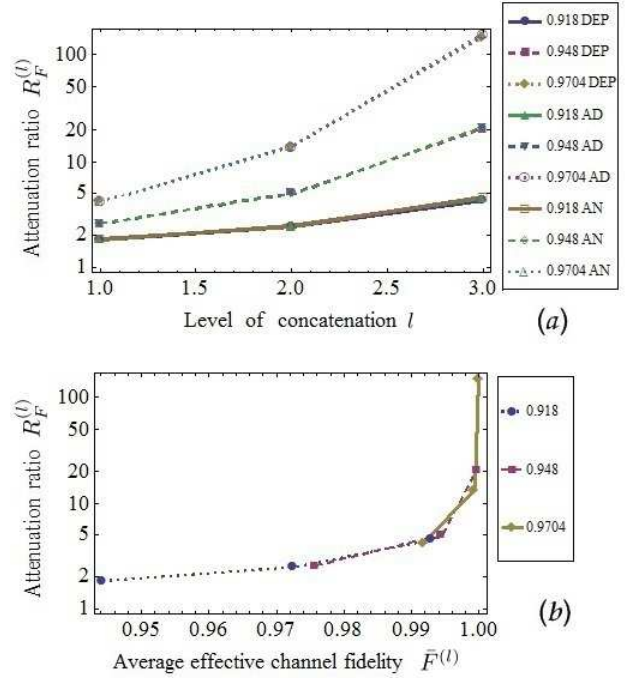}\\
\caption{(Color online) (a) Three typical cases are considered: (i) $F^{\mathrm{avg}(0)}=0.9704$, $f=0.98, k=0.02$, (ii) $F^{\mathrm{avg}(0)}=0.948$, $f=0.98, k=1/15$, and (iii) $F^{\mathrm{avg}(0)}=0.918$, $f=0.94, k=0.05$. For each case, three noise models (depolarizing noise, amplitude damping noise, and arbitrary generated numerical noise) are chosen for $\mathcal{N}$. After 3-level concatenated QEC is performed, numerical results of attenuation ratio of SD of channel fidelity in Eq.~(\ref{b8}) ($l=1,2,3$) are depicted. (b) $F^{\mathrm{avg}(0)}$ set as $0.9704, 0.948$, and $0.918$ respectively. After 3-level concatenated QEC is performed, the relationship between attenuation ratio of SD of channel fidelity in Eq.~(\ref{b8}) and the average effective channel fidelity in Eq.~(\ref{b6}) is depicted. In each typical case, three noise models are chosen for $\mathcal{N}$, and the curves coincided with each others. To have a better display, we only keep the curves when $\mathcal{N}$ is set as amplitude damping noise. }\label{figure6}
\end{figure}

(iii) Moreover, one can study the impact of average channel fidelity $F^{\mathrm{avg}(0)}=(1-k)f+0.5k$ on the fluctuation. First, the noise model of $\mathcal{N}$ is also chosen as amplitude damping noise, and then we set $F^{\mathrm{avg}(0)}=0.92,0.93,0.94,0.95,0.96$ respectively. For each fixed $F^{\mathrm{avg}(0)}$, $k$ increases from $0.01$ to $0.06$ with a step $0.01$, and $f$ can be obtained by $f=(F^{\mathrm{avg}(0)}-0.5k)/(1-k)$ (the value of $F^{\mathrm{avg}(0)}$ and $k$ should ensure $f\in[0,1]$). In each case, $N_{0}=50000$ unitary channels are generated independently and randomly ($F^{\mathrm{avg}(0)}\approx\bar{F}^{(0)}$, as $N_{0}=50000$), and then, one-level QEC are performed and numerical results can be obtained.

Numerical results for attenuation ratio of the SD of channel fidelity are shown in Fig.~\ref{figure5}. The results indicate that for fixed $F^{\mathrm{avg}(0)}$, attenuation ratio of SD of channel fidelity and attenuation ratios of SDs of diagonal elements of QPM almost have no change for different $f$ and $k$, and attenuation ratios of SDs of off-diagonal elements are increasing as $k$ increases. Meanwhile, attenuation ratios of all SDs (except the first column elements of QPM) are increasing with the increase of $F^{\mathrm{avg}(0)}$. In all cases, it indicates that attenuation ratio of SD of channel fidelity and attenuation ratios of SDs of diagonal elements of QPM are almost only dependent on the value of $F^{\mathrm{avg}(0)}$.

(iv) Finally, we study the impact of level $l$ and the average effective channel fidelity $\bar{F}^{(l)}$ on the fluctuation in concatenated QEC protocol. We consider three typical cases ($f=0.98, k=0.02; f=0.98, k=1/15; f=0.94, k=0.05$), and for each case we choose three different noise models (depolarizing noise, amplitude damping noise, and arbitrary generated numerical noise) for $\mathcal{N}$. In each case, $N_{0}=50000$ unitary channels are generated independently and randomly, and then 3-level QEC are performed and numerical results can be obtained.

\begin{figure}[tbph]
\centering
\includegraphics[width=0.50 \textwidth]{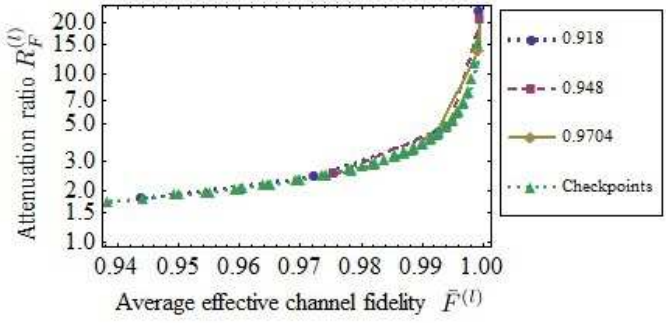}\\
\caption{(Color online) $\mathcal{N}$ is an amplitude damping channel, and the relationship between attenuation ratio of SD of channel fidelity in Eq.~(\ref{b8}) and the average effective channel fidelity in Eq.~(\ref{b6}) are depicted. A part of the data in this figure come from Fig.~\ref{figure4}(a) and Fig.~\ref{figure6}(b), and we also add another 42 points for one-level QEC, where $f$ is set as $0.9825,0.985,0.9875,0.99,0.9925,0.995$, and $0.9975$, respectively. For each $f$, the number $k$ increases from $0.01$ to $0.06$ with a step $0.01$).}\label{figure7}
\end{figure}

Numerical results for attenuation ratio of the SD of channel fidelity and the SD of channel fidelity are shown in Fig.~\ref{figure6}, and numerical results for the SDs of elements of QPM are obtained in Appendices~\ref{l3},~\ref{l4}, and~\ref{l5}. From the numerical results obtained, one can indicate that the noise model chosen for $\mathcal{N}$ has almost no impact on the fluctuation in each concatenated level. In first two levels, the SDs of off-diagonal elements of QPM decay more quickly (at least $80$ times faster) than those of diagonal elements (the minimum attenuation ratio of the SDs of diagonal elements is about 2.28), and the SDs of off-diagonal elements approach to $0$ when level $l$ increases to $2$. With the increase of level $l$, attenuation ratio of the SD of channel fidelity and attenuation ratios of the SDs of elements of QPM are increasing exponentially, and meanwhile the effective channels are approximate to depolarizing channel (after performing 2 levels concatenated QEC, the effective channels transform to a set of Pauli channels, and after one more level concatenated QEC, the effective channels transform to a set of depolarizing channels).

Moreover, as shown in Fig.~\ref{figure6}, attenuation ratio of the SD of channel fidelity is increasing exponentially with the increase of average effective channel fidelity $\bar{F}^{(l)}$ rather than the increase of level $l$. In order to test the relationship, we use the data of Fig.~\ref{figure4}(a) and other 42 points as checkpoints ($\mathcal{N}$ is set as amplitude damping noise, and $f=0.9825,0.985,0.9875,0.99,0.9925,0.995,0.9975$ respectively. For each $f$, $k$ increases from $0.01$ to $0.06$ with a step $0.01$, and one-level QEC is performed), and as shown in Fig.~\ref{figure7}, all checkpoints are almost situated on the curve Fig.~\ref{figure6}(b). These results indicate that attenuation ratio of the SD of channel fidelity is almost only dependent on the average effective channel fidelity $\bar{F}^{(l)}$ (note that in all cases considered in this work, attenuation ratio of the SD of channel fidelity has a significant linear correlation with attenuation ratios of the SDs of diagonal elements of QPM, and the correlation coefficient is about $0.8$).

In summary, based on the data shown in Fig.~\ref{figure4}(b), the relationship between SD of initial channel fidelity and proportionality constant $k$ could be
\be
\label{b9}
\delta F^{(0)}=0.354143k-0.0112724k^{2},
\ee
and based on the data shown in Fig.~\ref{figure7}, the relation between attenuation ratio of SD of channel fidelity and the average effective channel fidelity $\bar{F}^{(l)}$ (approximately equal to $F^{\mathrm{avg}(l)}$) could be
\be
\label{b10}
R^{(l)}_{F}&=&0.861795+\frac{0.300709}{\sqrt{1-\bar{F}^{(l)}}}\non \\
&\approx&0.861795+\frac{0.300709}{\sqrt{1-F^{\mathrm{avg}(l)}}}.
\ee
With Eq.~(\ref{b9}) and Eq.~(\ref{b10}), a rough estimation of the SD of channel fidelity in concatenated QEC protocol can be given. As an example, consider a case $f=0.99, k=0.04$, and the noise model of $\mathcal{N}$ is chosen as an arbitrarily generated numerical noise. After 3-level concatenated QEC is performed, the set of average channel fidelity defined in Eq.~(\ref{a17}) can be obtained,
\be
F^{\mathrm{avg}(0)}=0.9704, F^{\mathrm{avg}(1)}=0.991801,  \non\\
F^{\mathrm{avg}(2)}=0.99934, F^{\mathrm{avg}(3)}=0.999996. \non
\ee
Then, the SD of channel fidelity can be obtained according to
\be
\label{b11}
\delta F^{(l)}=\frac{\delta F^{(l-1)}}{R^{(l)}_{F}},
\ee
and the rough estimations of the SD of channel fidelity are obtained,
\be
\delta F^{(0)}&=&0.0141477,  \non\\
R^{(1)}_{F}=4.18277,\delta F^{(1)}&=&0.00338237, \non\\
R^{(2)}_{F}=12.5669,\delta F^{(2)}&=&0.00026915, \non\\
R^{(3)}_{F}=151.216, \delta F^{(3)}&=&1.7799\times10^{-6}.\non
\ee
Meanwhile, in numerical simulation one can obtain the set of average channel fidelity defined in Eq.~(\ref{b6}),
\be
\bar{F}^{(0)}=0.970407,\bar{F}^{(1)}=0.991808,\non\\
\bar{F}^{(2)}=0.999341,\bar{F}^{(3)}=0.999996 ,\non
\ee
and obtain the SD of channel fidelity,
\be
\delta F^{(0)}&=&0.0141059, \non\\
R^{(1)}_{F}=4.12663, \delta F^{(1)}&=&0.00341827, \non\\
R^{(2)}_{F}=13.9801,\delta F^{(2)}&=&0.000244509,  \non\\
R^{(3)}_{F}=156.274,\delta F^{(3)}&=&1.56462\times10^{-6}. \non
\ee
Now, one can indicate that the deviation between rough estimation and numerical simulation is increasing with the increase of concatenated level, because the rough estimation in next level is based on the value of the rough estimation in current level. There are 4 rough estimations when the concatenated level increases from 0 to 3, and the deviation between final rough estimation SD and final numerical simulation SD is about $13.8\%$, while the average deviation for each rough estimation is about $3.3\%$, which is acceptable. This example indicates that the relationships derived in Eq.~(\ref{b9}) and Eq.~(\ref{b10}) are convincible.

\begin{figure}[tbph]
\centering
\includegraphics[width=0.47 \textwidth]{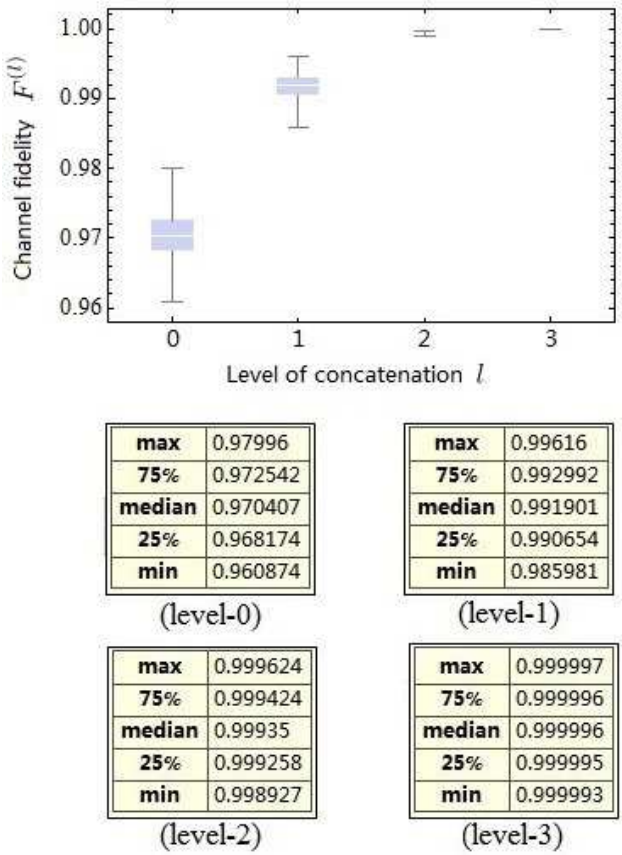}\\
\caption{(Color online) $\mathcal{N}$ is a depolarizing channel, and $f=0.98$, $k=0.02$. The fluctuation of channel fidelity of each concatenated level is depicted. }\label{figure13}
\end{figure}

In fact, the numerical calculations have been performed under the condition that the average initial channel fidelity is obviously above the error-correction threshold. Hence, there arise interesting questions whether the SD of channel fidelity has a threshold, and whether it is the same as the error-correction threshold. For 5-qubit code, the noise model of $\mathcal{N}$ and the value of $k$ have a little influence on the value of error-correction threshold. In numerical calculations, we choose depolarizing noise, amplitude damping noise, and arbitrary generated numerical noise for $\mathcal{N}$, and for each noise model of $\mathcal{N}$, $k$ increases from $0.01$ to $0.06$ with a step $0.01$. Then, we adjust the average initial channel fidelity $F^{\mathrm{avg}(0)}$ to ensure $F^{\mathrm{avg}(3)}\approx F^{\mathrm{avg}(2)}\approx F^{\mathrm{avg}(1)}\approx F^{\mathrm{avg}(0)}$ (absolutely equality cannot be reached except $\mathcal{N}$ is chosen as depolarizing noise). Finally, the numerical results of all different cases indicate that the values of attenuation ratio of SD of channel fidelity are almost the same, and the attenuation ratio of SD of channel fidelity in each concatenated level almost stabilize at $1.35$. In short, the attenuation ratio of SD of channel fidelity has a stable value about $1.35$ on the error-correction threshold.

After the main results of 5-qubit code, it is necessary to introduce the specific numerical simulations in detail. Because the numerical simulations are similar, we just give one specific numerical simulation as an example.

\begin{figure}[tbph]
\centering
\includegraphics[width=0.47 \textwidth]{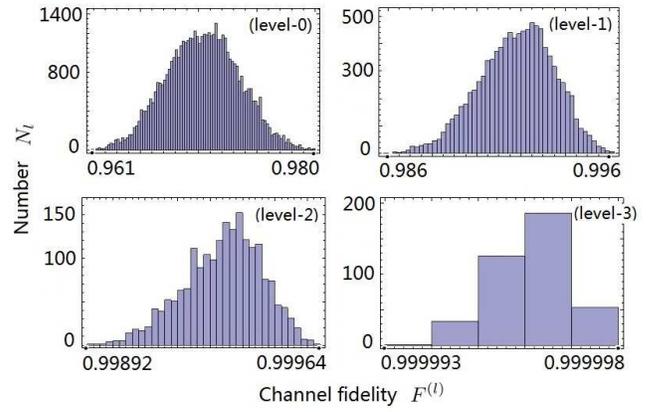}\\
\caption{(Color online) $\mathcal{N}$ is a depolarizing channel, and $f=0.98$, $k=0.02$. The exact distribution of channel fidelity of each concatenated level is depicted. }\label{figure14}
\end{figure}

Now, Let us consider a typical case, where the average initial channel fidelity $F^{\mathrm{avg}(0)}=0.9704$ (the channel fidelity $f=0.98$ for $\mathcal{N}$, and the proportionality constant $k=0.02$). There are 3 different noise models (depolarizing noise, amplitude damping noise, and arbitrary numerical noise) chosen for $\mathcal{N}$, and for each noise model, $N_{0}=50000$ unitary channels are generated independently and randomly. Here, we just discuss one case in detail, where $\mathcal{N}$ is set as a depolarizing noise channel. From the definition in Eq.~(\ref{b2}), one can obtain the average QPM of initial noise channels,
\be
\begin{tiny}\hat{\eta}^{\mathrm{avg}(0)}=\left(
\begin{array}{cccc}
 1 & 0 & 0 & 0 \\
 0 & 0.960533 & 0&0 \\
 0 & 0 & 0.960533 & 0 \\
 0 & 0 & 0 & 0.960533
\end{array}
\right),\non \end{tiny}
\ee
and then the concatenated 5-qubit QEC can be performed until the effective channel fidelity $F^{\mathrm{avg}(3)}\geq1-10^{-5}$. One can obtain the set of average channel fidelity defined in Eq.~(\ref{a17}),
\be
\label{e1}
F^{\mathrm{avg}(0)}&=&0.9704,F^{\mathrm{avg}(1)}=0.991801,\non\\
F^{\mathrm{avg}(2)}&=&0.99934,F^{\mathrm{avg}(3)}=0.999996.
\ee
Meanwhile, the set of average QPM $\eta^{\mathrm{avg}(l)}(l=0,1,2,3)$ are obtained in Eq. (\ref{s1}).

In numerical calculation, $N_{0}$ unitary channels are generated independently and randomly, and then $N_0$ samples of noise channel in Eq. (\ref{b1}) can be obtained. From Eq.~(\ref{b4}) one can obtain average QPM of the initial noise channels,
\be
\begin{tiny}\bar{\eta}^{(0)}=\left(
\begin{array}{cccc}
 1 & 0 & 0 & 0 \\
 0 & 0.960481 & 0.0000605251&0.0000418344 \\
 0 & -0.0000317435 & 0.960555 & 5.16027\times10^{-6} \\
 0 & 0.000078604 & -0.0000843357 & 0.960505
\end{array}
\right),\non \end{tiny}
\ee
and from Eq.~(\ref{b5}) we can obtain the SDs of elements of QPM of initial noise channels,
\be
\begin{tiny}\delta\eta^{(0)}=\left(
\begin{array}{cccc}
 0 & 0 & 0 & 0 \\
 0 &  0.0119545 & 0.0102918 & 0.0103645 \\
 0 & 0.0103015 &  0.0119311 & 0.0103342 \\
 0 & 0.0103549 & 0.0103438 &  0.0119042
\end{array}
\right).\non \end{tiny}
\ee
After performing 3-level concatenated QEC protocol constructed in Fig.~\ref{figure2}, the average QPMs $\bar{\eta}^{(l)}(l=0,1,2,3)$ and the SDs of elements of QPM $\delta\eta^{(l)}(l=0,1,2,3)$ are obtained in Eq. (\ref{s2}) and Eq. (\ref{s3}).

Meanwhile, in 3-level concatenated 5-qubit QEC, by Eq.~(\ref{b6}), one can obtain the average channel fidelity in each concatenated level,
\be
\label{e4}
\bar{F}^{(0)}=0.970385,\bar{F}^{(1)}=0.991792,\non\\
\bar{F}^{(2)}=0.999339,\bar{F}^{(3)}=0.999996,
\ee
and by Eq.~(\ref{b7}), one can obtain one set of SD of channel fidelity,
\be
\label{e5}
\delta F^{(0)}&=&0.00706234,\non\\ \delta F^{(1)}&=&0.00168564,\non\\
\delta F^{(2)}&=&0.000119716,\non\\ \delta F^{(3)}&=&8.24922\times10^{-7}.
\ee
The distribution of channel fidelity of each concatenated level are shown in Fig.~\ref{figure13} and Fig.~\ref{figure14}. The distribution of channel fidelity indicates that our numerical simulations are credible.

After introducing one specific numerical simulation, in the following we list the main numerical results for 5-qubit code.

When $F^{\mathrm{avg}(0)}=0.9704, f=0.98$, and $k=0.02$, we consider two cases where the noise model $\mathcal{N}$ is set as amplitude damping noise and randomly generated noise, respectively. For each case, $N_{0}=50000$ unitary channels are generated independently and randomly, and after performing 3-level concatenated 5-qubit QEC, the QPMs are obtained in Appendix~\ref{l3}. For the case $F^{\mathrm{avg}(0)}=0.9704, f=0.98, k=0.02$, the results of channel fidelity are obtained in Table~\ref{t1}.

\begin{table}
\caption{Results of 3-level concatenated QEC, where $F^{\mathrm{avg}(0)}=0.9704, f=0.98, k=0.02$, and $\mathcal{N}$ is set as depolarizing (DEP) noise, amplitude damping (AD) noise and arbitrary numerical (AN) noise, respectively.}\label{t1}
\centering\begin{tabular}{p{1.9cm} p{1.9cm} p{1.9cm} p{2.4cm} }
\hline\hline\noalign{\smallskip}
$\mathcal{N}-l$ & $F^{\mathrm{avg}(l)}$ & $\bar{F}^{(l)}$ & $\delta F^{(l)}$\\
\hline\noalign{\smallskip}
   DEP-$0$  & $0.9704$     & $0.970385$  &$0.00706234$ \\
   \hline\noalign{\smallskip}
   DEP-$1$  & $0.991801$    & $0.991792$   &$0.00168564$ \\
   \hline\noalign{\smallskip}
   DEP-$2$  & $0.99934$    & $0.999339$  & $0.000119716$  \\
    \hline\noalign{\smallskip}
   DEP-$3$  & $0.999996$   & $0.999996$   & $8.24922\times10^{-7}$  \\
\hline\noalign{\smallskip}
   AD-$0$  & $0.9704$     & $0.970367$  &$0.00706883$ \\
   \hline\noalign{\smallskip}
   AD-$1$  & $0.991803$    & $0.991785$   &$0.00171158$ \\
   \hline\noalign{\smallskip}
   AD-$2$  & $0.99934$    & $0.999337$  & $0.000124737$  \\
    \hline\noalign{\smallskip}
   AD-$3$  & $0.999996$   & $0.999996$   & $8.28466\times10^{-7}$  \\
\hline\noalign{\smallskip}
   AN-$0$  & $0.9704$     & $0.970393$  &$0.00707998$ \\
   \hline\noalign{\smallskip}
   AN-$1$  & $0.991801$    & $0.991795$   &$0.00171625$ \\
   \hline\noalign{\smallskip}
   AN-$2$  & $0.99934$    & $0.999339$  & $0.000122539$  \\
    \hline\noalign{\smallskip}
   AN-$3$  & $0.999996$   & $0.999996$   & $7.77169\times10^{-7}$  \\
\hline\hline
\end{tabular}
\end{table}

In addition, we also consider another two typical cases: $F^{\mathrm{avg}(0)}=0.948, f=0.98, k=1/15$ and $F^{\mathrm{avg}(0)}=0.918, f=0.94, k=0.05$. For each case, 3 different noise models (depolarizing noise, amplitude damping noise, and arbitrary numerical noise) are chosen for $\mathcal{N}$, and then $N_{0}=50000$ arbitrary unitary channels are generated independently and randomly. After 3-level concatenated 5-qubit QEC, the QPMs of the case ($F^{\mathrm{avg}(0)}=0.948, f=0.98, k=1/15$) are obtained in Appendix~\ref{l4}, and the results of channel fidelity are obtained in Table~\ref{t2}.

\begin{table}
\caption{Results of 3-level concatenated QEC, where $F^{\mathrm{avg}(0)}=0.948, f=0.98, k=1/15$, and $\mathcal{N}$ is set as depolarizing  (DEP) noise, amplitude damping (AD) noise and arbitrary numerical (AN) noise, respectively.}\label{t2}
\centering\begin{tabular}{p{1.9cm} p{1.9cm} p{1.9cm} p{2.4cm} }
\hline\hline\noalign{\smallskip}
$\mathcal{N}-l$ & $F^{\mathrm{avg}(l)}$ & $\bar{F}^{(l)}$ & $\delta F^{(l)}$\\
\hline\noalign{\smallskip}
   DEP-$0$  & $0.948$    & $0.948098$  & $0.0236263$   \\
    \hline\noalign{\smallskip}
   DEP-$1$  & $0.975956$    & $0.976057$  & $0.00923149$   \\
    \hline\noalign{\smallskip}
   DEP-$2$  & $0.994522$    & $0.994569$   & $0.00182526$  \\
    \hline\noalign{\smallskip}
   DEP-$3$  & $0.999704$    & $0.999708$  &  $0.0000898125$  \\

\hline\noalign{\smallskip}
   AD-$0$  & $0.948$    & $0.947807$  & $0.0236122$   \\
    \hline\noalign{\smallskip}
   AD-$1$  & $0.975958$    & $0.975832$  & $0.00923754$   \\
    \hline\noalign{\smallskip}
   AD-$2$  & $0.994523$    & $0.994462$   & $0.00188335$  \\
    \hline\noalign{\smallskip}
   AD-$3$  & $0.999704$    & $0.999697$  &  $0.0000893871$  \\

\hline\noalign{\smallskip}
    AN-$0$  & $0.948$    & $0.948126$  & $0.0235602$   \\
    \hline\noalign{\smallskip}
   AN-$1$  & $0.975956$    & $0.976061$  & $0.00917266$   \\
    \hline\noalign{\smallskip}
   AN-$2$  & $0.994522$    & $0.994565$   & $0.00185332$  \\
    \hline\noalign{\smallskip}
   AN-$3$  & $0.999704$    & $0.999708$  &  $0.0000881663$  \\

\hline\hline
\end{tabular}
\end{table}

For the case $F^{\mathrm{avg}(0)}=0.918, f=0.94, k=0.05$, the QPMs are obtained in Appendix~\ref{l5}, and the results of channel fidelity are obtained in Table~\ref{t3}.

\begin{table}
\caption{Results of the case ($F^{\mathrm{avg}(0)}=0.918, f=0.94, k=0.05$), $\mathcal{N}$ is set as depolarizing (DEP) noise, amplitude damping (AD) noise and arbitrary numerical (AN) noise respectively, and 3-level concatenated QEC protocol performed.}\label{t3}
\centering\begin{tabular}{p{1.9cm} p{1.9cm} p{1.9cm} p{2.4cm}}
\hline\hline\noalign{\smallskip}
$\mathcal{N}-l$ & $F^{\mathrm{avg}(l)}$ & $\bar{F}^{(l)}$ & $\delta F^{(l)}$\\
\hline\noalign{\smallskip}
    DEP-$0$  & $0.918$    & $0.917987$   & $0.0176449$  \\
    \hline\noalign{\smallskip}
   DEP-$1$  & $0.944226$    & $0.944207$   & $0.00969132$  \\
    \hline\noalign{\smallskip}
   DEP-$2$  & $0.972579$    & $0.97256$  & $0.00400429$   \\
   \hline\noalign{\smallskip}
   DEP-$3$  & $0.992929$    & $0.992916$  & $0.000939553$   \\

\hline\noalign{\smallskip}
   AD-$0$  & $0.918$    & $0.917932$   & $0.0176952$  \\
    \hline\noalign{\smallskip}
   AD-$1$  & $0.944263$    & $0.944193$   & $0.00970563$  \\
    \hline\noalign{\smallskip}
   AD-$2$  & $0.972613$    & $0.972552$  & $0.00394378$   \\
   \hline\noalign{\smallskip}
   AD-$3$  & $0.992946$    & $0.992915$  & $0.000886634$   \\

\hline\noalign{\smallskip}
    AN-$0$  & $0.918$    & $0.917966$   & $0.0176512$  \\
    \hline\noalign{\smallskip}
   AN-$1$  & $0.944226$    & $0.944208$   & $0.0096211$  \\
    \hline\noalign{\smallskip}
   AN-$2$  & $0.972579$    & $0.972555$  & $0.00404136$   \\
   \hline\noalign{\smallskip}
   AN-$3$  & $0.992929$    & $0.992918$  & $0.000892863$   \\

\hline\hline
\end{tabular}
\end{table}

For the three cases ($F^{\mathrm{avg}(0)}=0.9704, f=0.98, k=0.02, F^{\mathrm{avg}(0)}=0.948, f=0.98, k=1/15, F^{\mathrm{avg}(0)}=0.918, f=0.94, k=0.05$), the ranges of channel fidelity are shown in Table~\ref{t4}.

\begin{table}
\tiny
\caption{Channel fidelity ranges of the three cases, $\mathcal{N}$ is set as--depolarizing noise, amplitude damping noise, arbitrary numerical noise respectively.}\label{t4}
\centering\begin{tabular}{p{1.3cm} p{2cm} p{2cm} p{2.5cm} }
\hline\hline\noalign{\smallskip}

$Level-l$ & Depolarizing & Amplitude damping & Arbitrary numerical\\
 \hline\noalign{\smallskip}
   \multicolumn{3}{c}{$Case: F^{\mathrm{avg}(0)}=0.9704, f=0.98, k=0.02$ } \\
\hline\noalign{\smallskip}
   $3-0$  & $0.960874,0.97996$     & $0.960662,0.979986$  &$0.960959,0.980047$ \\
   \hline\noalign{\smallskip}
   $3-1$  & $0.985981,0.99616$    & $0.985836,0.996173$   &$0.986037,0.996193$ \\
   \hline\noalign{\smallskip}
   $3-2$  & $0.998927,0.999624$    & $0.998876,0.999662$  & $0.998829,0.999659$  \\
    \hline\noalign{\smallskip}
   $3-3$  & $0.999993,0.999997$   & $0.999992,0.999998$   & $0.999993,0.999997$  \\
 \hline\noalign{\smallskip}
   \multicolumn{3}{c}{$Case: F^{\mathrm{avg}(0)}=0.948, f=0.98, k=1/15$}  \\
    \hline\noalign{\smallskip}
   $3-0$  & $0.915798,0.980363$    & $0.916223,0.979665$  & $0.916544,0.980917$   \\
    \hline\noalign{\smallskip}
   $3-1$  & $0.941584,0.996312$    & $0.942071,0.996065$  & $0.942418,0.996511$   \\
    \hline\noalign{\smallskip}
   $3-2$  & $0.985955,0.998509$    & $0.981892,0.998827$   & $0.984649,0.998784$  \\
    \hline\noalign{\smallskip}
   $3-3$  & $0.999411,0.999897$    & $0.999323,0.999893$  &  $0.999299,0.999904$  \\
  \hline\noalign{\smallskip}
   \multicolumn{3}{c}{$Case: F^{\mathrm{avg}(0)}=0.918, f=0.94, k=0.05$}  \\
    \hline\noalign{\smallskip}
   $3-0$  & $0.894185,0.941837$    & $0.893903,0.942168$   & $0.893958,0.942072$  \\
    \hline\noalign{\smallskip}
   $3-1$  & $0.912102,0.970343$    & $0.91176,0.970697$   & $0.911864,0.970569$  \\
    \hline\noalign{\smallskip}
   $3-2$  & $0.956464,0.984939$    & $0.957988,0.98349$  & $0.955941,0.983458$   \\
   \hline\noalign{\smallskip}
   $3-3$  & $0.990172,0.995657$    & $0.990319,0.994978$  & $0.98965,0.995132$   \\
\hline\hline
\end{tabular}
\end{table}

\section{Main results of 7-qubit and 9-qubit code}
\label{vi}

In this section, we perform concatenated QEC with 7-qubit Steane code and 9-qubit Shor code for the noise in Eq.~(\ref{b1}), and the numerical simulations are obtained in subsections~\ref{via} and~\ref{vib}.

In the error correction with the Steane code, encoding process $\mathcal{V}$ is a unitary transformation $V$ in a $2^7$-dimensional Hilbert space, and its inverse $V^\dg$ is the decoding process. The set of correctable errors $\{E_{m}\}^{63}_{m=0}$ consists of the identity operator $E_{0}=\hat{I}^{\otimes7}_{2}$, all the weight-one Pauli operators $\sigma^{j}_{i} (i=1,2,3, j=1,2,...,7)$ and $42$ weight-two Pauli operators such as $\sigma^{1}_{1}\otimes\sigma^{2}_{3}, \sigma^{5}_{3}\otimes\sigma^{3}_{1},...$, etc. However, there are more than one choices for the correctable weight-two errors, and the typical one is the set of Pauli operators with exactly one $\sigma_{1}$ and one $\sigma_{3}$ error. In the error correction with the Shor code, encoding process $\mathcal{T}$ is a unitary transformation $T$ in a $2^9$-dimensional Hilbert space, and its inverse $T^\dg$ is the decoding process. The set of correctable errors $\{E_{m}\}^{255}_{m=0}$ consists of the identity operator $E_{0}=\hat{I}^{\otimes9}_{2}$, and other $255$ weight-one to weight-three Pauli operators.

From Sec.~\ref{iv}, for any quantum code, the average effective channel is dependent only on average of the initial noise channel. For the average noise channel in Eq. (\ref{b3}), when performing concatenated QEC with 5-qubit code, the average effective channel will be approximate to depolarizing channel quickly. When performing concatenated QEC with Steane code, the average effective channel will be approximate to a Pauli channel, and one can note that the weights of $\hat{\sigma}_1,\hat{\sigma}_2,\hat{\sigma}_3$ operators in this Pauli channel are dependent on the noise model of $\mathcal{N}$ and the proportionality constant $k$. When performing concatenated QEC with Shor code, the average effective channel will be approximate to one of Pauli-$X$ and Pauli-$Z$ channels, and in next concatenated level, the average effective channel will be approximate to the other. Meanwhile, for these codes, the degree of approximation increases with the concatenated level increases.

\subsection{Main results of 7-qubit Steane code}
\label{via}

\begin{figure}[tbph]
\centering
\includegraphics[width=0.46 \textwidth]{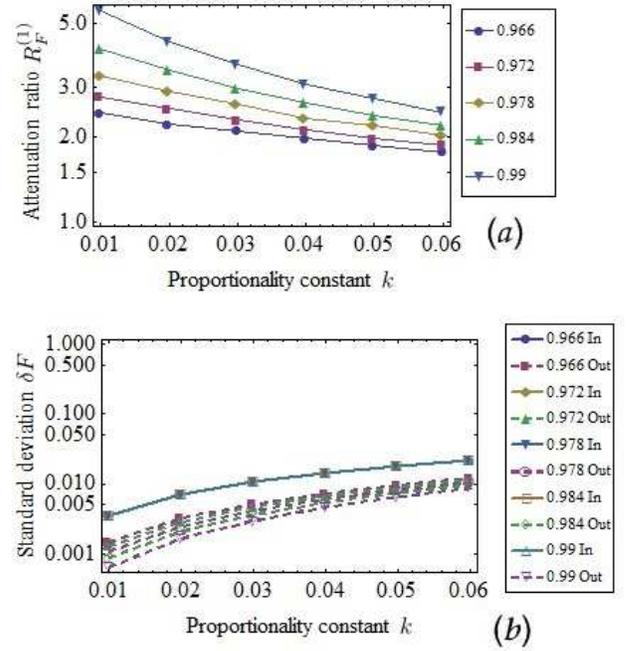}\\
\caption{(Color online) One-level QEC is performed, when $\mathcal{N}$ is set as amplitude damping noise: (a) Numerical results of attenuation ratio of SD of channel fidelity in Eq.~(\ref{b8}) ($l=1$) are depicted. The channel fidelity $f$ is set as $0.966, 0.972, 0.978, 0.984$, and $0.99$, respectively.
 (b) Numerical results of SD of channel fidelity in Eq.~(\ref{b7}) ($l=0,1$) are depicted. The channel fidelity $f$ is set as $0.966, 0.972, 0.978, 0.984$, and $0.99$, respectively. ``In'' means the case without performing QEC and ``Out'' means the case after performing one-level QEC.}\label{figure8}
\end{figure}

In our numerical simulations for Steane code, the SDs of elements of QPM have similar behavior as 5-qubit code: The SDs of diagonal elements of QPM have a significant linear correlation with SD of channel fidelity, and the SDs of off-diagonal elements of QPM decay more quickly (at least $50$ times faster) than those of diagonal elements. SDs of off-diagonal elements approach to 0 after 2 levels concatenated QEC. In the following, we mainly focus on the SD of channel fidelity.

For QEC with the Steane code, 4 variables in Eq. (\ref{b1}) are considered: the noise model of $\mathcal{N}$, $f$ (channel fidelity of $\mathcal{N}$), the proportionality constant $k$, and an arbitrary unitary channel $u(\bm\omega)$.

\begin{figure}[tbph]
\centering
\includegraphics[width=0.46 \textwidth]{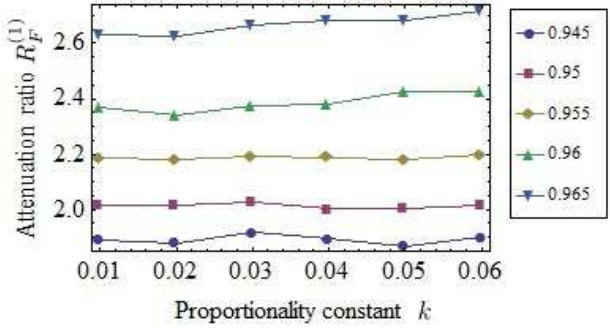}\\
\caption{(Color online) $\mathcal{N}$ is an amplitude damping channel, and $F^{\mathrm{avg}(0)}$ is set as $0.945,0.95,0.955,0.96,$ and $0.965$, respectively. For each $F^{\mathrm{avg}(0)}$, proportionality constant $k$ increases from $0.01$ to $0.06$ with a step $0.01$. After one-level QEC is performed, numerical results of attenuation ratio of SD of channel fidelity in Eq.~(\ref{b8}) ($l=1$) are depicted. }\label{figure9}
\end{figure}

(i) To study the impact of the noise model of $\mathcal{N}$ on the fluctuation, we set $f=0.98, k=0.02$, and the noise model of $\mathcal{N}$ is chosen as depolarizing noise, amplitude damping noise, and other 20 arbitrarily generated numerical noises. In each case $N_{0}=70000$ unitary channels are generated independently and randomly, and then one-level QEC are performed and numerical results are obtained.

Numerical results indicate that the noise model of $\mathcal{N}$ has nearly no influence on the SDs of channel fidelity. In all cases, the SD of channel fidelity decays with attenuation ratio $R^{(1)}_{F}\approx3.1$.

(ii) Next, we study the impacts of $f$ and $k$ on the fluctuation. The noise model of $\mathcal{N}$ is chosen as amplitude damping noise, and then we set $f=0.966,0.972,0.978,0.984,0.99$ respectively, and for each $f$, $k$ increases from $0.01$ to $0.06$ with a step $0.01$. In each case $N_{0}=70000$ unitary channels are generated independently and randomly, and then one-level QEC are performed and numerical results can be obtained.

Numerical results for the SD of channel fidelity are shown in Fig.~\ref{figure8}, and the results indicate that for initial noise channels, the SD of channel fidelity is almost only dependent on $k$. For a fixed $f$, attenuation ratios of SDs after QEC are decreasing as $k$ increases, while for a fixed $k$, attenuation ratios of SDs after QEC are increasing as $f$ increases.

(iii) Moreover, one can study the impact of average channel fidelity $F^{\mathrm{avg}(0)}=(1-k)f+0.5k$ on the fluctuation. Now, the noise model of $\mathcal{N}$ is also chosen as amplitude damping noise, and we set $F^{\mathrm{avg}(0)}=0.945,0.95,0.955,0.96,0.965$ respectively. For each fixed $F^{\mathrm{avg}(0)}$, $k$ increases from $0.01$ to $0.06$ with a step $0.01$, and $f$ can be obtained via $f=(F^{\mathrm{avg}(0)}-0.5k)/(1-k)$, where the values of $F^{\mathrm{avg}(0)}$ and $k$ should ensure $f\in[0,1]$. In each case, $N_{0}=70000$ unitary channels are generated independently and randomly ($F^{\mathrm{avg}(0)}\approx\bar{F}^{(0)}$, as $N_{0}=70000$), and then, one-level QEC are performed and numerical results can be obtained.

\begin{figure}[tbph]
\centering
\includegraphics[width=0.47 \textwidth]{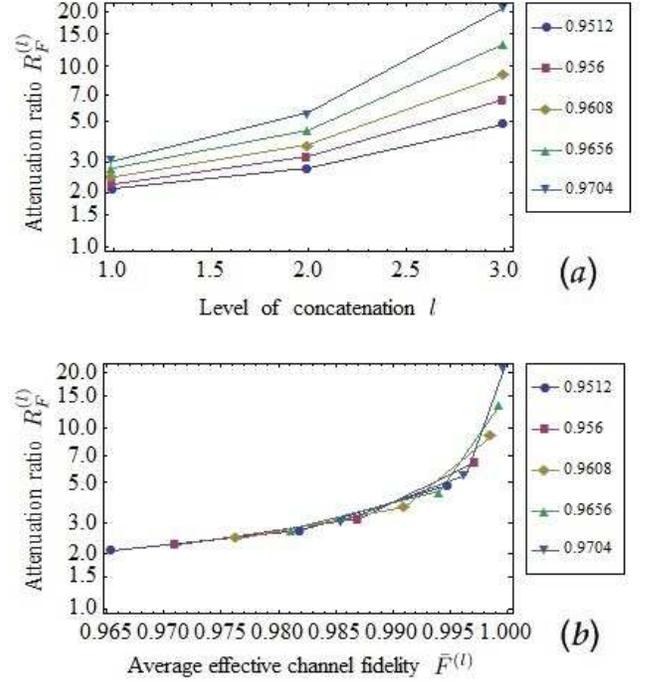}\\
\caption{(Color online) $\mathcal{N}$ is an amplitude damping channel, and five typical cases are considered: $f=0.98$, $k$ increases from $0.02$ to $0.06$ with a step $0.01$ (meanwhile, $F^{\mathrm{avg}(0)}$ equals $0.9704, 0.9656, 0.9608, 0.956$, and $0.9512$ respectively). After 3-level concatenated QEC is performed: (a) Numerical results of attenuation ratio of SD of channel fidelity in Eq.~(\ref{b8}) ($l=1,2,3$) are depicted. (b) The relationship between attenuation ratio of SD of channel fidelity in Eq.~(\ref{b8}) and the average effective channel fidelity in Eq.~(\ref{b6}) is depicted. }\label{figure10}
\end{figure}

Numerical results for attenuation ratio of the SD of channel fidelity are shown in Fig.~\ref{figure9}. The results indicate that for a fixed $F^{\mathrm{avg}(0)}$, attenuation ratio of SD of channel fidelity almost has the same value for different $f$ and $k$. Meanwhile, attenuation ratios of the SDs are increasing with the increase of $F^{\mathrm{avg}(0)}$. In all cases, it is indicated that attenuation ratio of SD of channel fidelity is almost dependent only on the value of $F^{\mathrm{avg}(0)}$.

(iv) Finally, we study the influences of level $l$ and the average effective channel fidelity $\bar{F}^{(l)}$ on the fluctuation in concatenated QEC protocol. We consider five typical cases, where $f=0.98$, and $k$ increases from $0.02$ to $0.06$ with a step $0.01$. In all these cases, the noise model of $\mathcal{N}$ is chosen as amplitude damping noise. In each case, $N_{0}=268912$ unitary channels are generated independently and randomly, and then 3-level QEC are performed and numerical results can be obtained.

\begin{figure}[tbph]
\centering
\includegraphics[width=0.50 \textwidth]{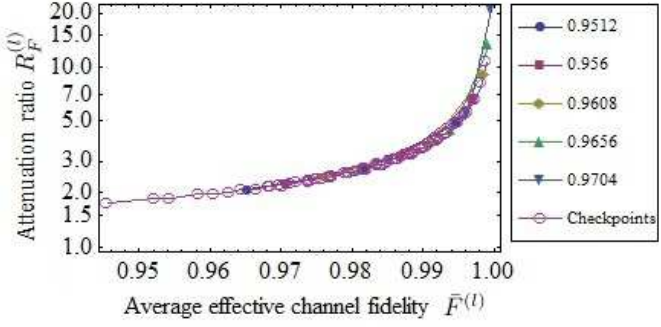}\\
\caption{(Color online) $\mathcal{N}$ is an amplitude damping channel, and the relationship between attenuation ratio of SD of channel fidelity in Eq.~(\ref{b8}) and the average effective channel fidelity in Eq.~(\ref{b6}) are depicted. A part of the data in this figure come from Fig.~\ref{figure8}(a) and Fig.~\ref{figure10}(b), and we also add another 42 points for one-level QEC, where $f$ is set as $0.9825,0.985,0.9875,0.99,0.9925,0.995$, and $0.9975$, respectively. For each $f$, the number $k$ increases from $0.01$ to $0.06$ with a step $0.01$).}\label{figure11}
\end{figure}

As shown in Fig.~\ref{figure10}, attenuation ratio of the SD of channel fidelity is increasing exponentially with the increase of average effective channel fidelity $\bar{F}^{(l)}$ rather than the increase of level $l$. In order to make this conclusion more reliable, we set $\mathcal{N}$ as an amplitude damping noise, $f=0.9825,0.985,0.9875,0.99,0.9925,0.995,0.9975$ respectively, where $k$ increases from $0.01$ to $0.06$ with a step $0.01$ for each $f$, and then one-level QEC is performed. From the numerical calculation, $42$ points can be obtained and together with the data in Fig.~\ref{figure8}(a), one can have $72$ checkpoints. As shown in in Fig.~\ref{figure11}, all checkpoints are almost situated on the curve in Fig.~\ref{figure10}(b). The results indicate that attenuation ratio of the SD of channel fidelity is almost only dependent on the average effective channel fidelity $\bar{F}^{(l)}$.

In addition, one can investigate the performance of channel fidelity's SD at the error-correction threshold for the Steane code. For the Steane code, the noise model of $\mathcal{N}$ and the value of $k$ have a little influence on the value of error-correction threshold. In numerical calculations, we choose depolarizing noise, amplitude damping noise, and arbitrary generated numerical noise for $\mathcal{N}$, and for each noise model of $\mathcal{N}$, $k$ increases from $0.01$ to $0.06$ with a step $0.01$. Then, we adjust the average initial channel fidelity $F^{\mathrm{avg}(0)}$ to ensure $F^{\mathrm{avg}(3)}\approx F^{\mathrm{avg}(2)}\approx F^{\mathrm{avg}(1)}\approx F^{\mathrm{avg}(0)}$. Finally, the numerical results of all different cases indicate that the values of attenuation ratio of SD of channel fidelity are almost the same, and the attenuation ratio of SD of channel fidelity in each concatenated level almost stabilize at $1.62$. In summary, the attenuation ratio of SD of channel fidelity has a stable value about $1.62$ on the error-correction threshold.

\subsection{Main results of 9-qubit Shor code}
\label{vib}

\begin{figure}[tbph]
\centering
\includegraphics[width=0.47 \textwidth]{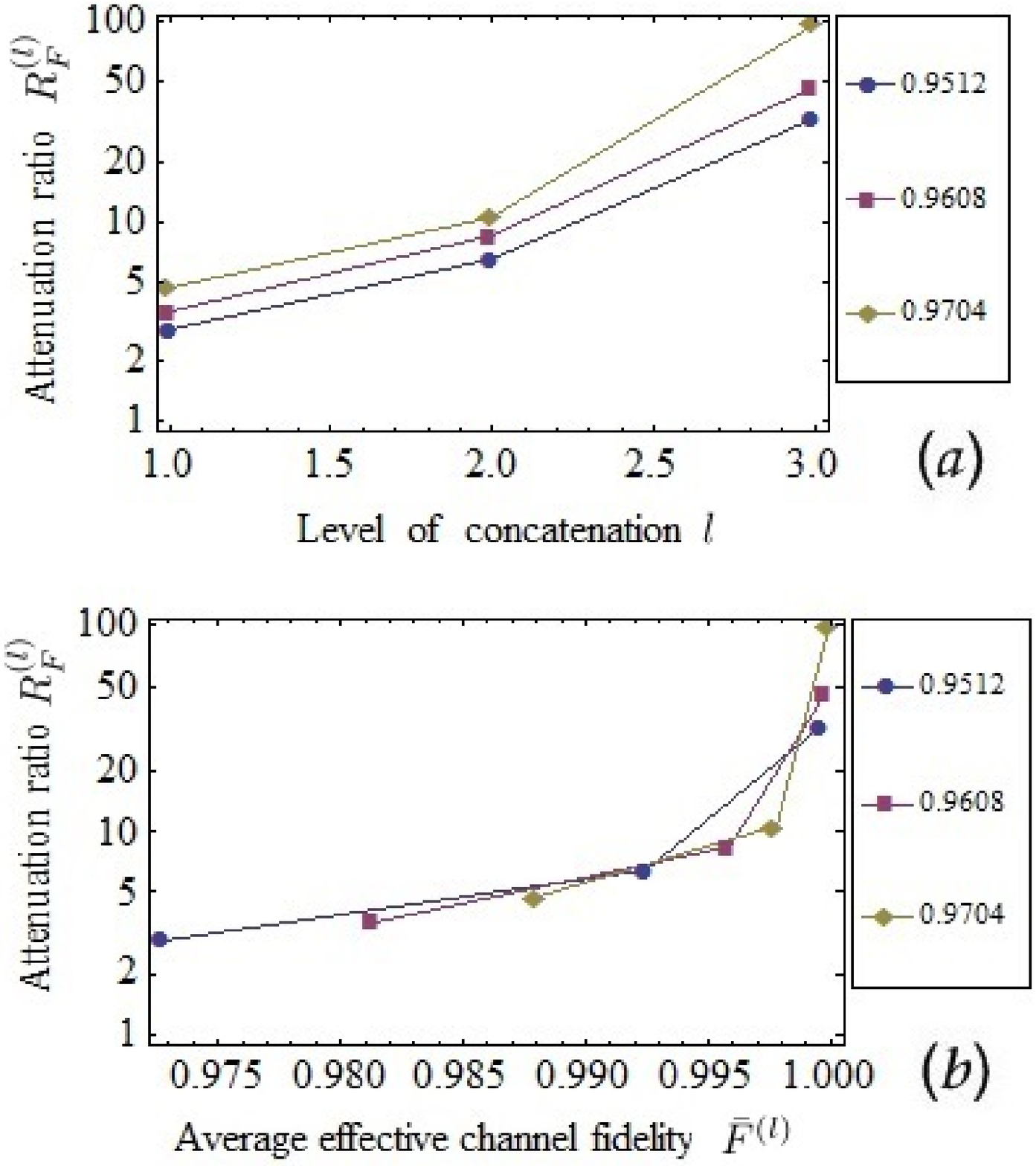}\\
\caption{(Color online) $\mathcal{N}$ is an amplitude damping channel, and three typical cases are considered: $f=0.98$, $k$ increases from $0.02$ to $0.06$ with a step $0.02$ (meanwhile, $F^{\mathrm{avg}(0)}$ equals $0.9704, 0.9608$, and $0.9512$ respectively). After 3-level concatenated QEC is performed: (a) Numerical results of attenuation ratio of SD of channel fidelity in Eq.~(\ref{b8}) ($l=1,2,3$) are depicted. (b) The relationship between attenuation ratio of SD of channel fidelity in Eq.~(\ref{b8}) and the average effective channel fidelity in Eq.~(\ref{b6}) is depicted.}\label{figure12}
\end{figure}

In our numerical simulations for Shor code, there are similar behaviors with 5-qubit code and Steane code. The SDs of $\eta^{(l)}_{22}$ in QPM have a significant linear correlation with SD of channel fidelity, and the SDs of off-diagonal elements of QPM decay more quickly (at least $30$ times faster) than those of diagonal elements. SDs of off-diagonal elements approach to 0 after 2 levels concatenated QEC. However, there are also different behaviors from 5-qubit code or Steane code. The noise model of $\mathcal{N}$ in Eq.~(\ref{b1}) has obvious influence on attenuation ratios of SDs of channel fidelity. The effective channels are approximate to one of Pauli-$X$ and Pauli-$Z$ channels in each level, and in next concatenated level, the effective channels are approximate to the other. Meanwhile, the degree of approximation increases with the concatenated level increases. Therefore, in the following, we mainly focus on the performance of SDs of channel fidelity and diagonal elements in QPM.

(i) To study the impact of the noise model of $\mathcal{N}$ on the fluctuation, one can set $f=0.98, k=0.02$, and the noise model of $\mathcal{N}$ is chosen as depolarizing noise, amplitude damping noise, and other 20 arbitrarily generated numerical noises. In each case $N_{0}=90000$ unitary channels are generated independently and randomly, and then one-level QEC are performed and numerical results are obtained. Numerical results indicate that the noise model of $\mathcal{N}$ has obvious influence on attenuation ratios of SDs of channel fidelity. In all cases, the SDs of channel fidelity decay with attenuation ratio about $3.4\sim4.7$.

(ii) To clear the relationship between attenuation ratios of SD of channel fidelity and average effective channel fidelity, the noise model of $\mathcal{N}$ is fixed as amplitude damping noise, and we set $f=0.98, k=0.02,0.04,0.06$, respectively. In each case, $N_{0}=72900$ unitary channels are generated independently and randomly, and then three-level QEC are performed and numerical results are obtained. As shown in Fig.~\ref{figure12}, numerical results indicate that attenuation ratio of SD of channel fidelity is almost only dependent on the average effective channel fidelity $\bar{F}^{(l)}$.

(iii) In our numerical simulations, it indicates that the effective channels are approximate to one of Pauli-$X$ and Pauli-$Z$ channels in each level, and in next concatenated level, the effective channels are approximate to the other. Meanwhile, the degree of approximation increases with the concatenated level increases, and the fluctuation of noise channel decays exponentially as concatenated QEC performed.

For the two cases in Appendix~\ref{l6}, the weight of $\hat{\sigma}_2$ decays rapidly, and the relative weights of $\hat{\sigma}_1$ and $\hat{\sigma}_3$ are oscillating in adjacent concatenated levels. Meanwhile, SDs of $\eta^{(l)}_{11}$ and $\eta^{(l)}_{33}$ are oscillating and decaying in adjacent concatenated levels. Moreover, other arbitrarily generated numerical noise models of $\mathcal{N}$ are also simulated, where for each fixed $\mathcal{N}$, $f=0.98, k=0.02,0.04,0.06$, respectively.  The numerical results show that the SDs of diagonal elements in QPM have similar behaviors.

In addition, one can investigate the performance of channel fidelity's SD at the error-correction threshold for Shor code. When performing QEC for the average of physical noise channel in Eq.~(\ref{b3}), the noise model of $\mathcal{N}$ and the value of $k$ have obvious influence on the value of error-correction threshold. Therefore, for different $\mathcal{N}$ and $k$, we adjust the average initial channel fidelity $F^{\mathrm{avg}(0)}$ to ensure $F^{\mathrm{avg}(3)}\approx F^{\mathrm{avg}(2)}\approx F^{\mathrm{avg}(1)}\approx F^{\mathrm{avg}(0)}$. In numerical calculations, we choose depolarizing noise, amplitude damping noise, and arbitrary generated numerical noise for $\mathcal{N}$, and for each noise model of $\mathcal{N}$, $k$ increases from $0.02$ to $0.06$ with a step $0.02$. Finally, the numerical results of all different cases indicate that the values of attenuation ratio of SD of channel fidelity are roughly the same, and the attenuation ratio of SD of channel fidelity in each concatenated level roughly stabilize at $2$. In summary, the attenuation ratio of SD of channel fidelity has a stable value about $2$ around the error-correction threshold.

\section{The analysis of numerical results}
\label{vii}

Lots of numerical results have been obtained in Sec.~\ref{v} and Sec.~\ref{vi}, and in order to test the robustness of concatenated QEC protocol, different $\mathcal{N}$, $f$ and $k$ in Eq.~(\ref{b1}) have been chosen. As shown in the two sections, 5-qubit code is more robust than 7-qubit and 9-qubit code. More specifically, the initial SDs of channel fidelity and SDs of elements in QPM are determined by $k$, and the SDs decreases exponentially with consecutive levels of concatenation. When $\mathcal{N}$, $f$ and $k$ are fixed, it seems that the attenuation ratio of SD of channel fidelity has linear relationship with the attenuation ratio of average error rate. However, for different codes, the specific processes of decreasing of fluctuation have a little difference. Finally, the numerical results indicate that there is no threshold below which the standard deviation does not decrease.

In order to understand the numerical results, further analysis is necessary. For noise channel $\varepsilon$, the error rate $r$ is
\be
\label{f1}
r=1-F,
\ee
and here, $F$ is the channel fidelity of $\varepsilon$, then $\delta r=\delta F$. Now, for noise channel $\varepsilon$, there always are relationship between the element $\eta_{\mu\nu} (\mu,\nu=0,1,2,3)$ of QPM and $r$,
\be
\label{f2}
\eta_{\mu\nu}(r)=C_{\mu\nu}(r),
\ee
where $C_{\mu\nu}$ is a function depends on noise model. If $\eta_{\mu\nu}(r)$ expressed with Taylor expansion
\be
\label{f3}
\eta_{\mu\nu}(r)=\sum_{n=0}^{n}c_{\mu\nu,n}r^{n},
\ee
here, $c_{\mu\nu,n}$ is the coefficient of the $n$-th order $r$ ($\mu,\nu=0,1,2,3$), and it is dependent on the specific noise model of $\varepsilon$. When the noise channels belong to the same distribution of one independent error model, the coefficient $c_{\mu\nu,n}$ can be viewed fixed,
\be
\label{f10}
\delta\eta_{\mu\nu}(r)&=&\sqrt{\frac{1}{N}\sum_{N=1}^{N}(\eta_{\mu\nu,N}-\eta_{\mu\nu}^{\mathrm{avg}})^{2}} \non \\
&=&\sqrt{\frac{1}{N}\sum_{N=1}^{N}(\sum_{n=0}^{n}c_{\mu\nu,n}r_{N}^{n}-\sum_{n=0}^{n}c_{\mu\nu,n}(r^{\mathrm{avg}})^{n})^{2}} \non \\
&=&\sqrt{\sum_{n=0}^{n}c_{\mu\nu,n}^{2}\frac{1}{N}\sum_{N=1}^{N}(r_{N}^{n}-(r^{\mathrm{avg}})^{n})^{2}} \non \\
&=&\sqrt{\sum_{n=0}^{n}(c_{\mu\nu,n}\delta r^{n})^{2}}=\sqrt{\sum_{n=1}^{n}(c_{\mu\nu,n}\delta r^{n})^{2}},
\ee
here, $N$ is the number of the noise channels of the same distribution. It is indicated in Eq.~(\ref{f10}) that SDs of elements of QPM have  approximate linear relationship with SD of error rate when $r$ is small.

Similarly with Eq.~(\ref{b3}), the average QPM of $l$-th level effective channel can be expressed as
\begin{equation}
\label{f5}
{\eta}^{\mathrm{avg}(l)}=(1-k^{(l)})\eta^{(l)}_{\alpha}+k^{(l)}\eta^{(l)}_{\beta},
\end{equation}
where $k^{(l)}$ is a proportionality coefficient, $\eta^{(l)}_{\alpha}$ is the QPM of fixed noise model, and $\eta^{(l)}_{\beta}$ is the average QPM of noise with fluctuation.

Then, the numerical results can be divided into three cases. (i) In the case when average channel fidelity set above error-correction threshold, after $l$-level concatenated QEC, the average error rate of effective channels $r^{\mathrm{avg}(l)}$ approaches $0$. So, the error rate of every effective channel approaches $0$, either. The effective QPMs will approach the constant matrix
\begin{equation}
\label{f4}
r^{(l)}\rightarrow0,
{\eta}^{(l)}\rightarrow\left(\begin{array}{cccc}
1 & 0& 0& 0\\
0 & 1& 0& 0\\
0 & 0& 1& 0\\
0 & 0& 0& 1\\\end{array}
\right),
\end{equation}
and based on Eq.~(\ref{f10}), it is expected that the fluctuation of noise channels would approach $0$ and attenuation ratios of the SDs have approximate linear correlation with attenuation ratio of error rate.

For 5-qubit code, when choosing different specific noise model, average error rate is almost the same in each concatenated level, as is the SD of channel fidelity. For 7-qubit code and 9-qubit code, when choosing different specific noise model, average error rate in each concatenated level is different and dependent on the initial noise model, and so is the SD of channel fidelity. In this case, average error rate can be used as an important index to judge the fluctuation. Meanwhile, 5-qubit code is more robust than 7-qubit and 9-qubit codes when against different noise.

(ii) In another case when average channel fidelity set below error-correction threshold, after $l$-level concatenated QEC in our numerical simulations, the effective QPMs will approach the constant matrix
\begin{equation}
\label{f9}
r^{(l)}\rightarrow0.75,
{\eta}^{(l)}\rightarrow\left(\begin{array}{cccc}
1 & 0& 0& 0\\
0 & 0& 0& 0\\
0 & 0& 0& 0\\
0 & 0& 0& 0\\\end{array}
\right),
\end{equation}
and based on Eq.~(\ref{f10}), one may think that the coefficient $c_{\mu\nu,n}$ approaches $0$ is the reason why the fluctuation is decreased in noise channels.

(iii) In the third case when average channel fidelity equals error-correction threshold, there is a typical case for 5-qubit code where $\eta^{(0)}_{\alpha}$ $\eta^{(0)}_{\beta}$ are set as depolarizing noise. In this case,
\begin{equation}
\label{f6}
{\eta}^{\mathrm{avg}(l)}={\eta}^{\mathrm{avg}(l-1)}={\eta}^{\mathrm{avg}(0)},
\end{equation}
so,
\begin{equation}
\label{f7}
{\eta}^{\mathrm{avg}(l)}={\eta}^{\mathrm{avg}(0)}
=\left(\begin{array}{cccc}
1 & 0& 0& 0\\
0 & \frac{\sqrt{6}}{3}& 0& 0\\
0 & 0& \frac{\sqrt{6}}{3}& 0\\
0 & 0& 0& \frac{\sqrt{6}}{3}\\\end{array}\right),
\end{equation}
and the numerical results indicate the attenuation ratio of SD of channel fidelity has a stable value for different $k^{(0)}$. So,
it is expected that $k^{(l)}$ decreases with consecutive levels of concatenation. Because diagonal elements of $\eta^{(0)}_{\alpha}$ are larger than ${\eta}^{\mathrm{avg}(0)}$'s, and diagonal elements of $\eta^{(0)}_{\beta}$ are smaller than ${\eta}^{\mathrm{avg}(0)}$'s, when $k^{(l)}$ decreases, it is required that $\eta^{(l)}_{\alpha}$ and $\eta^{(l)}_{\beta}$ approach ${\eta}^{\mathrm{avg}(l)}$ with consecutive levels of concatenation to ensure Eq.~(\ref{f5}) and Eq.~(\ref{f7}),
\begin{equation}
\label{f8}
\eta^{(0)}_{\alpha,\mu\mu}=\frac{4f-1}{3}>{\eta}^{\mathrm{avg}(0)}_{\mu\mu},\\
\eta^{(0)}_{\beta,\mu\mu}=\frac{1}{3}<{\eta}^{\mathrm{avg}(0)}_{\mu\mu},\mu=1,2,3.
\end{equation}
Based on the results of 5-qubit code, we guess 5-qubit code had the ability to make both $\eta^{(l)}_{\alpha}$ and $\eta^{(l)}_{\beta}$ approach ${\eta}^{\mathrm{avg}(l)}$, and then result in decreasing of the SD on error-correction threshold. In addition, we guess it is similar for 7-qubit code and 9-qubit code, with different details.

In summary, we conjecture that the ability of error correct codes concentrate effective channels to the fixed channel is the reason decreasing of fluctuation.

\section{Remarks and discussion}
\label{viii}

In this work, we consider fluctuation because the noise in QEC is always approximately known, and the estimations for noise process in experiments indicate that fluctuation exists in physical noise channel. Therefore, the physical noise channel may be appropriately described by the average noise channel and fluctuation. So, it is necessary to consider the general and realistic case that each noise channel in every qubit is generated randomly and independently from one distribution. In the previous work~\cite{Ball}, robustness of stabilizer codes for mixed channels was discussed, and for arbitrary noise models with fixed channel fidelity, the efficient of QEC with 5-qubit code was discussed in Ref.~\cite{Huang2}. However, the fluctuation of noise channels was not considered in these two works, and in the present work, the robustness of concatenated QEC protocol against noise with fluctuation is studied. On the other hand, the authors in Ref.~\cite{C.C2017} introduced the robustness of hard decoding optimization algorithm against noise with perturbations, and it is indicated that the fixed concatenated QEC protocol is also robust to noise with fluctuation in our work.

For the cases we have considered, the numerical results indicate that concatenated QEC protocols with 5-qubit code, Steane code and Shor code are efficient and robust to noise with fluctuation. More specifically, SD of channel fidelity and SDs of diagonal elements of QPM always decay exponentially with the increase of concatenated level. Meanwhile, SDs of off-diagonal elements of QPM decay more quickly than those of diagonal elements, and SDs of off-diagonal elements approach to 0 after 2 levels concatenated QEC. For 5-qubit code and Steane code, different noise model of physical noise almost has no influence on the attenuation ratios of the SDs of channel fidelity, which are almost only dependent on the average effective channel fidelity. For Shor code, different noise model of physical noise has obvious influence on the attenuation ratios of the SDs of channel fidelity. For 5-qubit code, the effective channels are approximate to depolarizing channel as the concatenated level increases. For Steane code, the effective channels are approximate to one Pauli channel as the concatenated level increases. For Shor code, the effective channels are approximate to one of Pauli-$X$ and Pauli-$Z$ channels, and in the next concatenated level, the effective channels are approximate to the other. Moreover, for these three codes, the degree of approximation increases with the concatenated level increases. On the error-correction threshold, the attenuation ratio of SD of channel fidelity roughly has a stable value, for 5-qubit code it is about $1.35$, for Steane code it is about $1.62$, and for Shor code it is about $2$.

In the recent works~\cite{M.Guti,S. J. Beale,D. Greenbaum,E. Huang}, it is indicated that the effective noise channels approach Pauli-like channels and the off-diagonal elements of QPM disappear in the concatenated QEC protocol. In the work~\cite{Huang}, the effective noise channels are approximate to depolarizing channel as the concatenated 5-qubit QEC performed. In the work~\cite{S. J. Beale}, off-diagonal elements of an error process matrix decay more quickly than the diagonals in a quantum error correcting code. In this paper, numerical calculations show that the results of these works are also valid for average channel when considering fluctuation, and in Sec.~\ref{iv}, it is proved that when considering fluctuation of physical noise channel, the average effective channel is dependent only on the average of physical noise channel. Note that the average of physical noise channel here plays the role of the independent error model in the previous works, and now, one may conclude that in the independent error model, the results in previous works are also valid for average channel where fluctuation exists. Our numerical simulations with fluctuating noise do confirm this conjecture. Meanwhile, the numerical simulations also show that, in the concatenated QEC, one can suppress physical noise channels with fluctuation, and obtain effective channel with higher fidelity and smaller fluctuation. In addition, we believe this conjecture is valid in general, not limited to the noise in Eq.~(\ref{b1}).

\acknowledgments
This work was supported by the National Natural Science Foundation of China under Grants No.~11405136 and No.~11847307, and the Fundamental Research Funds for the Central Universities under Grant No.~2682019LK11.

\appendix
\begin{widetext}
\section{QPM for the case when average initial channel fidelity is 0.9704}
\label{l3}

(i) Consider the case where $F^{\mathrm{avg}(0)}=0.9704$ ($f=0.98, k=0.02$), the noise model of $\mathcal{N}$ is set as a depolarizing noise. Now, the initial noise channels is a case as in Eq. (\ref{b3}). With QPT in 3-level concatenated 5-qubit QEC protocol, and from the definition in Eq.~(\ref{a14}), the average QPM in each concatenated level ($l=0,1,2,3$) can be obtained,
\be
\label{s1}
\hat{\eta}^{\mathrm{avg}(0)}=
\left(
\begin{array}{cccc}
 1 & 0 & 0 & 0 \\
 0 & 0.960533 & 0&0 \\
 0 & 0 & 0.960533 & 0 \\
 0 & 0 & 0 & 0.960533
\end{array}
\right) ,
\hat{\eta}^{\mathrm{avg}(1)}=\left(
\begin{array}{cccc}
 1 & 0 & 0 & 0 \\
 0 & 0.989068 & 0&0 \\
 0 & 0 & 0.989068 & 0 \\
 0 & 0 & 0 & 0.989068
\end{array}
\right) ,\non\\
\hat{\eta}^{\mathrm{avg}(2)}=\left(
\begin{array}{cccc}
 1 & 0 & 0 & 0 \\
 0 & 0.99912 & 0 & 0 \\
 0 & 0 & 0.99912 & 0 \\
 0 & 0 & 0 & 0.99912
\end{array}
\right) ,
\hat{\eta}^{\mathrm{avg}(3)}=\left(
\begin{array}{cccc}
 1 & 0 & 0 & 0 \\
 0 & 0.999994 & 0 & 0 \\
 0 & 0 & 0.999994 & 0 \\
 0 & 0 & 0 & 0.999994
\end{array}
\right).
\ee
Meanwhile, in numerical calculation, $N_{0}=50000$ unitary channels are generated independently and randomly, and then $N_{0}=50000$ samples of noise channel in Eq. (\ref{b1}) can be obtained. With QPT in 3-level concatenated 5-qubit QEC protocol, and from the definition in Eq. (\ref{b4}), the average QPM in each concatenated level ($l=0,1,2,3$) can be obtained,
\be
\label{s2}
&\bar{\eta}^{(0)}=
\left(
\begin{array}{cccc}
 1 & 0 & 0 & 0 \\
 0 & 0.960481 & 0.0000605251&0.0000418344 \\
 0 & -0.0000317435 & 0.960555 & 5.16027\times10^{-6} \\
 0 & 0.000078604 & -0.0000843357 & 0.960505
\end{array}
\right),\non\\
&\bar{\eta}^{(1)}=\left(
\begin{array}{cccc}
  1 & 0 & 0 & 0 \\
 0 & 0.989053 & 1.14253\times10^{-8} & -2.57205\times10^{-8} \\
 0 & 3.13037\times10^{-8} & 0.989057 & 2.24322\times10^{-8} \\
 0 & 8.5616\times10^{-9} & -3.69532\times10^{-9} & 0.989057
\end{array}
\right),\non\\
&\bar{\eta}^{(2)}=\left(
\begin{array}{cccc}
 1 & 0 & 0 & 0 \\
 0 & 0.999118 & 0 & 0 \\
 0 & 0 & 0.999118 & 0 \\
 0 & 0 & 0 & 0.999118
\end{array}
\right),
\bar{\eta}^{(3)}=\left(
\begin{array}{cccc}
 1 & 0 & 0 & 0 \\
 0 & 0.999994 & 0 & 0 \\
 0 & 0 & 0.999994 & 0 \\
 0 & 0 & 0 & 0.999994
\end{array}
\right) .
\ee
Moreover, from the definition in Eq. (\ref{b5}), the SDs of elements of QPM in each concatenated level can be obtained,
\be
\label{s3}
\delta\eta^{(0)}=
\left(
\begin{array}{cccc}
 0 & 0 & 0 & 0 \\
 0 & 0.0119545 & 0.0102918 & 0.0103645 \\
 0 & 0.0103015 & 0.0119311 & 0.0103342 \\
 0 & 0.0103549 & 0.0103438 & 0.0119042
\end{array}
\right),
\delta\eta^{(1)}=\left(
\begin{array}{cccc}
 0 & 0 & 0 & 0 \\
 0 &0.00225611 & 1.6228\times10^{-6} & 1.61845\times10^{-6} \\
 0 & 1.6426\times10^{-6} &0.0022524 & 1.61305\times10^{-6} \\
 0 & 1.61442\times10^{-6} & 1.6371\times10^{-6} & 0.00225269
\end{array}
\right),\non\\
\delta\eta^{(2)}=\left(
\begin{array}{cccc}
 0 & 0 & 0 & 0 \\
 0 & 0.000159632 & 0 & 0 \\
 0 & 0 &0.000159636 & 0 \\
 0 & 0 & 0 &0.000159644
\end{array}
\right),
\delta\eta^{(3)}=\left(
\begin{array}{cccc}
 0 & 0 & 0 & 0 \\
 0 & 1.0298\times10^{-6} & 0 & 0 \\
 0 & 0 &1.0298\times10^{-6} & 0 \\
 0 & 0 & 0 &1.0298\times10^{-6}
\end{array}
\right).\non\\
\ee

(ii) Consider the case where $F^{\mathrm{avg}(0)}=0.9704$ ($f=0.98, k=0.02$), and the noise model of $\mathcal{N}$ is set as amplitude damping noise. Similar to Eq. (\ref{s1}), one can obtain,
\be
&\eta^{\mathrm{avg}(0)}=
\left(
\begin{array}{cccc}
 1 & 0 & 0 & 0 \\
 0 & 0.966968 & 0 & 0 \\
 0 & 0 & 0.966968 & 0 \\
 0.039002 & 0 & 0 & 0.947665
\end{array}
\right),
\eta^{\mathrm{avg}(1)}=\left(
\begin{array}{cccc}
 1 & 0 & 0 & 0 \\
 0 & 0.989356 & 0 & 0 \\
 0 & 0 & 0.989356 & 0 \\
 -2.25619\times10^{-8} & 0 & 0 & 0.988501
\end{array}
\right),\non\\
&\eta^{\mathrm{avg}(2)}=\left(
\begin{array}{cccc}
 1 & 0 & 0 & 0 \\
 0 & 0.999121 & 0 & 0 \\
 0 & 0 & 0.999121 & 0 \\
 0 & 0 & 0 & 0.999119
\end{array}
\right),
\eta^{\mathrm{avg}(3)}=\left(
\begin{array}{cccc}
 1 & 0 & 0 & 0 \\
 0 & 0.999994 & 0 & 0 \\
 0 & 0 & 0.999994 & 0 \\
 0 & 0 & 0 & 0.999994
\end{array}
\right) .\non
\ee
Meanwhile, similar to Eq. (\ref{s2}), one can obtain,
\be
&\bar{\eta}^{(0)}=
\left(
\begin{array}{cccc}
 1 & 0 & 0 & 0 \\
 0 & 0.96693 & 3.92608\times10^{-6} & -0.0000258762 \\
 0 & 0.0000231165 & 0.966933 & 0.0000427746 \\
 0.039002 & -0.000044454 & -0.0000644307 & 0.947603
\end{array}
\right),\non\\
&\bar{\eta}^{(1)}=\left(
\begin{array}{cccc}
 1 & 0 & 0 & 0 \\
 5.79578\times10^{-9} & 0.989335 & -1.3483\times10^{-8} & 1.61674\times10^{-7} \\
 -2.33067\times10^{-8} & -1.73751\times10^{-9} & 0.989333 & -1.30353\times10^{-7} \\
 -9.14585\times10^{-8} & -9.4923\times10^{-8} & 8.72004\times10^{-9} & 0.988474
\end{array}
\right),\non\\
&\bar{\eta}^{(2)}=\left(
\begin{array}{cccc}
 1 & 0 & 0 & 0 \\
 0 & 0.999117 & 0 & 0 \\
 0 & 0 & 0.999117 & 0 \\
 0 & 0 & 0 & 0.999115
\end{array}
\right),
\bar{\eta}^{(3)}=\left(
\begin{array}{cccc}
 1 & 0 & 0 & 0 \\
 0 & 0.999994 & 0 & 0 \\
 0 & 0 & 0.999994 & 0 \\
 0 & 0 & 0 & 0.999994
\end{array}
\right) .\non
\ee
Similar to Eq. (\ref{s3}), one can obtain,
\be
&\delta\eta^{(0)}=
\left(
\begin{array}{cccc}
 0 & 0 & 0 & 0 \\
 0 & 0.0119593 & 0.0103135 & 0.0103279 \\
 0 & 0.0103322 & 0.0119385 & 0.010331 \\
 0 & 0.0103092 & 0.0103496 & 0.0119574
\end{array}
\right),\non\\
&\delta\eta^{(1)}=\left(
\begin{array}{cccc}
 0 & 0 & 0 & 0 \\
 4.51878\times10^{-6} &0.00230821 & 1.65042\times10^{-6} & 8.47241\times10^{-6} \\
 4.40404\times10^{-6} & 1.65422\times10^{-6} & 0.00230987 & 8.51862\times10^{-6} \\
 4.48766\times10^{-6} & 8.12115\times10^{-6} & 8.23066\times10^{-6} & 0.00230203
\end{array}
\right),\non\\
&\delta\eta^{(2)}=\left(
\begin{array}{cccc}
 0 & 0 & 0 & 0 \\
 0 & 0.000166349 & 0 & 0 \\
 0 & 0 &0.000166318 & 0 \\
 0 & 0 & 0 &0.000166307
\end{array}
\right),
\delta\eta^{(3)}=\left(
\begin{array}{cccc}
 0 & 0 & 0 & 0 \\
 0 & 1.05564\times10^{-6} & 0 & 0 \\
 0 & 0 &1.05564\times10^{-6} & 0 \\
 0 & 0 & 0 &1.05564\times10^{-6}
\end{array}
\right) .\non
\ee

(iii) Consider the case where $F^{\mathrm{avg}(0)}=0.9704$ ($f=0.98, k=0.02$), and the noise model of $\mathcal{N}$ is set as arbitrary numerical noise. The set of Kraus operators $\{A_{0},A_{1},A_{2},A_{3}\}$ are generated randomly as
\be
&A_{0}=\left(
\begin{array}{cc}
 0.756784 & -0.0493575+0.0480098 i \\
 -0.0493575-0.0480098 i & 0.78349
\end{array}
\right),
A_{1}=\left(
\begin{array}{cc}
 0.0267779 -0.0260467 i & -0.00125374+0.0452789 i \\
 0.0308079 & -0.0267779+0.0260467 i
\end{array}
\right),\non\\
&A_{2}=\left(
\begin{array}{cc}
 0.0349194 +0.0339659 i & 0.0401748 \\
 -0.00163493-0.0590456 i & -0.0349194-0.0339659 i
\end{array}
\right),
A_{3}=\left(
\begin{array}{cc}
 0.638225 & 0.0599647 -0.0583273 i \\
 0.0599647 +0.0583273 i & 0.605781
\end{array}
\right).\non
\ee
Similar to Eq. (\ref{s1}), one can obtain,
\be
&\eta^{\mathrm{avg}(0)}=
\left(
\begin{array}{cccc}
 1 & 0 & 0 & 0 \\
 -0.0027981 & 0.963156 & 0.00780802 & 0.00217159 \\
 -0.00272169 & 0.00780802 & 0.962724 & 0.00211229 \\
 -0.000756966 & 0.00217159 & 0.00211229 & 0.955716
\end{array}
\right),\non\\
&\eta^{\mathrm{avg}(1)}=\left(
\begin{array}{cccc}
 1 & 0 & 0 & 0 \\
 5.66186\times10^{-7} & 0.9891 & -7.71093\times10^{-10} &-2.18793\times10^{-7} \\
 5.56773\times10^{-7} & -8.85474\times10^{-10} & 0.989115 & -2.10952\times10^{-7} \\
 4.56715\times10^{-7} & -2.18543\times10^{-7} & -2.10678\times10^{-7} & 0.988988
\end{array}
\right),\non\\
&\eta^{\mathrm{avg}(2)}=\left(
\begin{array}{cccc}
 1 & 0 & 0 & 0 \\
 0 & 0.99912 & 0 & 0 \\
 0 & 0 & 0.99912 & 0 \\
 0 & 0 & 0 & 0.99912
\end{array}
\right),
\eta^{\mathrm{avg}(3)}=\left(
\begin{array}{cccc}
 1 & 0 & 0 & 0 \\
 0 & 0.999994 & 0 & 0 \\
 0 & 0 & 0.999994 & 0 \\
 0 & 0 & 0 & 0.999994
\end{array}
\right) .\non
\ee
Meanwhile, similar to Eq. (\ref{s2}), one can obtain,
\be
&\bar{\eta}^{(0)}=
\left(
\begin{array}{cccc}
 1 & 0 & 0 & 0 \\
 -0.0027981 & 0.963166 & 0.00782901 & 0.00217924 \\
 -0.00272169 & 0.00780882 & 0.962688 & 0.00201543 \\
 -0.000756966 & 0.00217252 & 0.00210951 & 0.955716
\end{array}
\right),\non\\
&\bar{\eta}^{(1)}=\left(
\begin{array}{cccc}
 1 & 0 & 0 & 0 \\
 5.65903\times10^{-7} & 0.989093 & -3.14165\times10^{-8} & -2.30899\times10^{-7} \\
 5.62561\times10^{-7} & -1.15879\times10^{-8} & 0.989102 & -2.14612\times10^{-7} \\
 4.62737\times10^{-7} &-2.10859\times10^{-7} & -2.17221\times10^{-7} & 0.988986
\end{array}
\right),\non\\
&\bar{\eta}^{(2)}=\left(
\begin{array}{cccc}
 1 & 0 & 0 & 0 \\
 0 & 0.999119 & 0 & 0 \\
 0 & 0 & 0.999119 & 0 \\
 0 & 0 & 0 & 0.999119
\end{array}
\right),
\bar{\eta}^{(3)}=\left(
\begin{array}{cccc}
 1 & 0 & 0 & 0 \\
 0 & 0.999994 & 0 & 0 \\
 0 & 0 & 0.999994 & 0 \\
 0 & 0 & 0 & 0.999994
\end{array}
\right) .\non
\ee
Moreover, similar to Eq. (\ref{s3}), one can obtain,
\be
&\delta\eta^{(0)}=
\left(
\begin{array}{cccc}
 0 & 0 & 0 & 0 \\
 0 & 0.0119532 & 0.0103064 & 0.0103111 \\
 0 & 0.0103195 & 0.0119472 & 0.0103344 \\
 0 & 0.0102979 & 0.0103475 & 0.011935
\end{array}
\right),\non\\
&\delta\eta^{(1)}=\left(
\begin{array}{cccc}
 0 & 0 & 0 & 0 \\
 5.86592\times10^{-7} &0.00229476 & 2.18785\times10^{-6} & 2.24794\times10^{-6} \\
 5.78341\times10^{-7} & 2.18117\times10^{-6} & 0.00230084 & 2.29403\times10^{-6} \\
 6.43318\times10^{-7} & 2.26602\times10^{-6} & 2.25332\times10^{-6} & 0.00229437
\end{array}
\right),\non\\
&\delta\eta^{(2)}=\left(
\begin{array}{cccc}
 0 & 0 & 0 & 0 \\
 0 & 0.000163399 & 0 & 0 \\
 0 & 0 &0.000163398 & 0 \\
 0 & 0 & 0 &0.000163387
\end{array}
\right),
\delta\eta^{(3)}=\left(
\begin{array}{cccc}
 0 & 0 & 0 & 0 \\
 0 & 1.02346\times10^{-6} & 0 & 0 \\
 0 & 0 &1.02346\times10^{-6} & 0 \\
 0 & 0 & 0 &1.02346\times10^{-6}
\end{array}
\right) .\non
\ee

\section{QPM for the case when average initial channel fidelity equals 0.948}
\label{l4}

(i) Consider the case where $F^{\mathrm{avg}(0)}=0.948$ ($f=0.98, k=1/15$), and the noise model of $\mathcal{N}$ is set as depolarizing noise. Similar to Eq. (\ref{s1}), one can obtain,
\be
&\eta^{\mathrm{avg}(0)}=
\left(
\begin{array}{cccc}
 1 & 0 & 0 & 0 \\
 0 & 0.930667 & 0&0 \\
 0 & 0 & 0.930667 & 0 \\
 0 & 0 & 0 & 0.930667
\end{array}
\right),
\eta^{\mathrm{avg}(1)}=\left(
\begin{array}{cccc}
 1 & 0 & 0 & 0 \\
 0 & 0.967942 & 0&0 \\
 0 & 0 & 0.967942 & 0 \\
 0 & 0 & 0 & 0.967942
\end{array}
\right),\non\\
&\eta^{\mathrm{avg}(2)}=\left(
\begin{array}{cccc}
 1 & 0 & 0 & 0 \\
 0 & 0.992696 & 0 & 0 \\
 0 & 0 & 0.992696 & 0 \\
 0 & 0 & 0 & 0.992696
\end{array}
\right),
\eta^{\mathrm{avg}(3)}=\left(
\begin{array}{cccc}
 1 & 0 & 0 & 0 \\
 0 & 0.999605 & 0 & 0 \\
 0 & 0 & 0.999605 & 0 \\
 0 & 0 & 0 & 0.999605
\end{array}
\right) .\non
\ee
Meanwhile, similar to Eq. (\ref{s2}), one can obtain,
\be
&\bar{\eta}^{(0)}=
\left(
\begin{array}{cccc}
 1 & 0 & 0 & 0 \\
 0 & 0.93102 & 0.000221614&0.000193132 \\
 0 &-0.0000653772 & 0.930593 & -0.0000294828 \\
 0 & -0.0000187266 & 0.000180216 & 0.930778
\end{array}
\right),\non\\
&\bar{\eta}^{(1)}=\left(
\begin{array}{cccc}
  1 & 0 & 0 & 0 \\
 0 & 0.968083 & 4.90894\times10^{-7} & -7.15038\times10^{-7} \\
 0 & 1.85508\times10^{-8} & 0.968058 & -2.12155\times10^{-7} \\
 0 & 1.18343\times10^{-6} & -1.83959\times10^{-7} & 0.968088
\end{array}
\right),\non\\
&\bar{\eta}^{(2)}=\left(
\begin{array}{cccc}
 1 & 0 & 0 & 0 \\
 0 & 0.992758 & 0 & 0 \\
 0 & 0 & 0.992758 & 0 \\
 0 & 0 & 0 & 0.992758
\end{array}
\right),
\bar{\eta}^{(3)}=\left(
\begin{array}{cccc}
 1 & 0 & 0 & 0 \\
 0 & 0.999611 & 0 & 0 \\
 0 & 0 & 0.999611 & 0 \\
 0 & 0 & 0 & 0.999611
\end{array}
\right) .\non
\ee
Moreover, similar to Eq. (\ref{s3}), one can obtain,
\be
\delta\eta^{(0)}=
\left(
\begin{array}{cccc}
 0 & 0 & 0 & 0 \\
 0 & 0.0397003 & 0.0343901 & 0.0342946 \\
 0 & 0.0343729 & 0.0398946 & 0.0343628 \\
 0 & 0.0343119 & 0.0343456 & 0.0398591
\end{array}
\right),
\delta\eta^{(1)}=\left(
\begin{array}{cccc}
 0 & 0 & 0 & 0 \\
 0 &0.0124017& 0.0000572596 & 0.0000580205 \\
 0 &0.0000568981 & 0.0124559 & 0.0000582792 \\
 0 & 0.0000575347 & 0.000057466 & 0.0124273
\end{array}
\right),\non\\
\delta\eta^{(2)}=\left(
\begin{array}{cccc}
 0 & 0 & 0 & 0 \\
 0 & 0.00243353 & 0 & 0 \\
 0 & 0 &0.00243383 & 0 \\
 0 & 0 & 0 &0.0024337
\end{array}
\right),
\delta\eta^{(3)}=\left(
\begin{array}{cccc}
 0 & 0 & 0 & 0 \\
 0 & 0.000119704 & 0 & 0 \\
 0 & 0 &0.000119704 & 0 \\
 0 & 0 & 0 &0.000119704
\end{array}
\right) .\non
\ee

(ii) Consider the case where $F^{\mathrm{avg}(0)}=0.948$ ($f=0.98, k=1/15$), and the noise model of $\mathcal{N}$ is set as amplitude damping noise. Similar to Eq. (\ref{s1}), one can obtain,
\be
&\eta^{\mathrm{avg}(0)}=
\left(
\begin{array}{cccc}
 1 & 0 & 0 & 0 \\
 0 & 0.936795 & 0 & 0 \\
 0 & 0 & 0.936795 & 0 \\
 0.0371448 & 0 & 0 & 0.918411
\end{array}
\right),
\eta^{\mathrm{avg}(1)}=\left(
\begin{array}{cccc}
 1 & 0 & 0 & 0 \\
 0 & 0.968187& 0 & 0 \\
 0 & 0 & 0.968187 & 0 \\
 -1.76778\times10^{-8} & 0 & 0 & 0.967458
\end{array}
\right),\non\\
&\eta^{\mathrm{avg}(2)}=\left(
\begin{array}{cccc}
 1 & 0 & 0 & 0 \\
 0 & 0.992697 & 0 & 0 \\
 0 & 0 & 0.992697 & 0 \\
 0 & 0 & 0 & 0.992696
\end{array}
\right),
\eta^{\mathrm{avg}(3)}=\left(
\begin{array}{cccc}
 1 & 0 & 0 & 0 \\
 0 & 0.999605 & 0 & 0 \\
 0 & 0 & 0.999605 & 0 \\
 0 & 0 & 0 & 0.999605
\end{array}
\right) .\non
\ee
Meanwhile, similar to Eq. (\ref{s2}), one can obtain,
\be
&\bar{\eta}^{(0)}=
\left(
\begin{array}{cccc}
 1 & 0 & 0 & 0 \\
 0 & 0.936585 & 0.0000710796 & -0.00015667 \\
 0 & -0.0000462149 & 0.936586 & 0.000262721 \\
 0.0371448 & 0.000226651 & -0.000278215 & 0.918057
\end{array}
\right),\non\\
&\bar{\eta}^{(1)}=\left(
\begin{array}{cccc}
 1 & 0 & 0 & 0 \\
 -2.64726\times10^{-7} & 0.968004 & 1.39364\times10^{-7} & -6.0835\times10^{-7} \\
 1.51358\times10^{-7} & 6.72634\times10^{-8} & 0.968008 & 1.10081\times10^{-6} \\
 5.51091\times10^{-7} & 3.39781\times10^{-7} & -1.0577\times10^{-6} & 0.967316
\end{array}
\right),\non\\
&\bar{\eta}^{(2)}=\left(
\begin{array}{cccc}
 1 & 0 & 0 & 0 \\
 0 & 0.992616 & 0 & 0 \\
 0 & 0 & 0.992616 & 0 \\
 0 & 0 & 0 & 0.992615
\end{array}
\right),
\bar{\eta}^{(3)}=\left(
\begin{array}{cccc}
 1 & 0 & 0 & 0 \\
 0 & 0.999597 & 0 & 0 \\
 0 & 0 & 0.999597 & 0 \\
 0 & 0 & 0 & 0.999597
\end{array}
\right) .\non
\ee
Moreover, similar to Eq. (\ref{s3}), one can obtain
\be
&\delta\eta^{(0)}=
\left(
\begin{array}{cccc}
 0 & 0 & 0 & 0 \\
 0 & 0.0398053 & 0.034424 & 0.0345028 \\
 0 & 0.0344167 & 0.039859 & 0.0344479 \\
 0 & 0.0345101 & 0.0344406 & 0.0398652
\end{array}
\right),
\delta\eta^{(1)}=\left(
\begin{array}{cccc}
 0 & 0 & 0 & 0 \\
 0.0000443976 &0.0124843 & 0.0000582924 & 0.0000629405 \\
 0.0000447759 & 0.0000575945 & 0.0124882 & 0.000062752 \\
 0.0000445216 & 0.0000618792 & 0.0000623155 & 0.0124369
\end{array}
\right),\non\\
&\delta\eta^{(2)}=\left(
\begin{array}{cccc}
 0 & 0 & 0 & 0 \\
 0 & 0.00251103 & 0 & 0 \\
 0 & 0 &0.00251094 & 0 \\
 0 & 0 & 0 &0.00251149
\end{array}
\right),
\delta\eta^{(3)}=\left(
\begin{array}{cccc}
 0 & 0 & 0 & 0 \\
 0 & 0.000119213 & 0 & 0 \\
 0 & 0 &0.000119213 & 0 \\
 0 & 0 & 0 &0.00011922
\end{array}
\right) .\non
\ee

(iii) Consider the case where $F^{\mathrm{avg}(0)}=0.948$ ($f=0.98, k=1/15$), and the noise model of $\mathcal{N}$ is set as an arbitrary numerical noise. The set of Kraus operators $\{A_{0},A_{1},A_{2},A_{3}\}$ are generated randomly as
\be
&A_{0}=\left(
\begin{array}{cc}
 -0.524991 & 0.0326596 +0.0688087 i \\
 0.0326596 -0.0688087 i & -0.41504
\end{array}
\right),
A_{1}=\left(
\begin{array}{cc}
 -0.0197012-0.0415073 i & 0.0567946 -0.0695926 i \\
 0.0235008 & 0.0197012 +0.0415073 i
\end{array}
\right),\non\\
&A_{2}=\left(
\begin{array}{cc}
 0.0120814 -0.0254537 i & -0.0144115 \\
 -0.0348284-0.0426766 i & -0.0120814+0.0254537 i
\end{array}
\right),
A_{3}=\left(
\begin{array}{cc}
 0.842943 & 0.0168194 +0.0354358 i \\
 0.0168194 -0.0354358 i & 0.899567
\end{array}
\right).\non
\ee
Similar to Eq. (\ref{s1}), one can obtain,
\be
&\eta^{\mathrm{avg}(0)}=
\left(
\begin{array}{cccc}
 1 & 0 & 0 & 0 \\
 -0.00260038 & 0.927991 & -0.00320695 & -0.00256223 \\
 0.00547861 & -0.00320695 & 0.933226 & 0.00539822 \\
 0.00437719 & -0.00256223 & 0.00539822 & 0.930782
\end{array}
\right),\non\\
&\eta^{\mathrm{avg}(1)}=\left(
\begin{array}{cccc}
 1 & 0 & 0 & 0 \\
 9.11371\times10^{-7} & 0.967925 & -1.65265\times10^{-7} &-2.03343\times10^{-7} \\
 -9.56857\times10^{-7} & -1.65205\times10^{-7} & 0.967929 & 2.64089\times10^{-7} \\
 -9.2898\times10^{-7} & -2.02772\times10^{-7} & 2.62992\times10^{-7} & 0.967972
\end{array}
\right),\non\\
&\eta^{\mathrm{avg}(2)}=\left(
\begin{array}{cccc}
 1 & 0 & 0 & 0 \\
 0 & 0.992696 & 0 & 0 \\
 0 & 0 & 0.992696 & 0 \\
 0 & 0 & 0 & 0.992696
\end{array}
\right),
\eta^{\mathrm{avg}(3)}=\left(
\begin{array}{cccc}
 1 & 0 & 0 & 0 \\
 0 & 0.999605 & 0 & 0 \\
 0 & 0 & 0.999605 & 0 \\
 0 & 0 & 0 & 0.999605
\end{array}
\right) .\non
\ee
Meanwhile, similar to Eq. (\ref{s2}), one can obtain,
\be
&\bar{\eta}^{(0)}=
\left(
\begin{array}{cccc}
 1 & 0 & 0 & 0 \\
 -0.00260038 & 0.928045 & -0.00308538 & -0.00240647 \\
 0.00547861 & -0.00324119 & 0.933364 & 0.00538546 \\
 0.00437719 & -0.00266455 & 0.00516673 & 0.931096
\end{array}
\right),\non\\
&\bar{\eta}^{(1)}=\left(
\begin{array}{cccc}
 1 & 0 & 0 & 0 \\
 8.82276\times10^{-7} & 0.96808 & -6.73593\times10^{-7} & 3.20803\times10^{-7} \\
 -9.60698\times10^{-7} &6.91649\times10^{-7} & 0.968072 & 9.9162\times10^{-7} \\
 -8.23116\times10^{-7} &-3.22704\times10^{-7} & 1.90669\times10^{-7} & 0.968091
\end{array}
\right),\non\\
&\bar{\eta}^{(2)}=\left(
\begin{array}{cccc}
 1 & 0 & 0 & 0 \\
 0 & 0.992753 & 0 & 0 \\
 0 & 0 & 0.992753 & 0 \\
 0 & 0 & 0 & 0.992753
\end{array}
\right),
\bar{\eta}^{(3)}=\left(
\begin{array}{cccc}
 1 & 0 & 0 & 0 \\
 0 & 0.999611 & 0 & 0 \\
 0 & 0 & 0.999611 & 0 \\
 0 & 0 & 0 & 0.999611
\end{array}
\right) .\non
\ee
Moreover, similar to Eq. (\ref{s3}), one can obtain,
\be
&\delta\eta^{(0)}=
\left(
\begin{array}{cccc}
 0 & 0 & 0 & 0 \\
 0 & 0.0397408 & 0.0343724 & 0.0344593 \\
 0 & 0.0344591 & 0.0397235 & 0.0343378 \\
 0 & 0.0343726 & 0.0344246 & 0.039625
\end{array}
\right),
\delta\eta^{(1)}=\left(
\begin{array}{cccc}
 0 & 0 & 0 & 0 \\
 9.07779\times10^{-6} &0.0123322 & 0.0000586173 & 0.0000587624 \\
 9.09903\times10^{-6} & 0.0000587095 & 0.0123587 & 0.0000598963 \\
 9.09537\times10^{-6} & 0.0000583646 & 0.0000587631 & 0.0123706
\end{array}
\right),\non\\
&\delta\eta^{(2)}=\left(
\begin{array}{cccc}
 0 & 0 & 0 & 0 \\
 0 & 0.00247113 & 0 & 0 \\
 0 & 0 &0.002471 & 0 \\
 0 & 0 & 0 &0.00247116
\end{array}
\right),
\delta\eta^{(3)}=\left(
\begin{array}{cccc}
 0 & 0 & 0 & 0 \\
 0 & 0.000117587 & 0 & 0 \\
 0 & 0 &0.000117587 & 0 \\
 0 & 0 & 0 &0.000117586
\end{array}
\right) .\non
\ee

\section{QPM for the case when average initial channel fidelity equals 0.918}
\label{l5}

(i) Consider the case where $F^{\mathrm{avg}(0)}=0.918$ ($f=0.94, k=0.05$), and the noise model of $\mathcal{N}$ is set as a depolarizing noise. Similar to Eq. (\ref{s1}), one can obtain,
\be
&\eta^{\mathrm{avg}(0)}=
\left(
\begin{array}{cccc}
 1 & 0 & 0 & 0 \\
 0 & 0.890667 & 0&0 \\
 0 & 0 & 0.890667 & 0 \\
 0 & 0 & 0 & 0.890667
\end{array}
\right),
\eta^{\mathrm{avg}(1)}=\left(
\begin{array}{cccc}
 1 & 0 & 0 & 0 \\
 0 & 0.925635 & 0&0 \\
 0 & 0 & 0.925635 & 0 \\
 0 & 0 & 0 & 0.925635
\end{array}
\right),\non\\
&\eta^{\mathrm{avg}(2)}=\left(
\begin{array}{cccc}
 1 & 0 & 0 & 0 \\
 0 & 0.963439 & 0 & 0 \\
 0 & 0 & 0.963439 & 0 \\
 0 & 0 & 0 & 0.963439
\end{array}
\right),
\eta^{\mathrm{avg}(3)}=\left(
\begin{array}{cccc}
 1 & 0 & 0 & 0 \\
 0 & 0.990572 & 0 & 0 \\
 0 & 0 & 0.990572 & 0 \\
 0 & 0 & 0 & 0.990572
\end{array}
\right) .\non
\ee
Meanwhile, similar to Eq. (\ref{s2}), one can obtain,
\be
&\bar{\eta}^{(0)}=
\left(
\begin{array}{cccc}
 1 & 0 & 0 & 0 \\
 0 & 0.890553 & 0.000080485&0.0000668321 \\
 0 & -0.0000760708 & 0.890553 & 0.000266967 \\
 0 & 0.000203702 & 9.53205\times10^{-7} & 0.890841
\end{array}
\right),\non\\
&\bar{\eta}^{(1)}=\left(
\begin{array}{cccc}
  1 & 0 & 0 & 0 \\
 0 & 0.925605 & -1.56384\times10^{-7} & 2.93121\times10^{-7} \\
 0 & 3.04631\times10^{-8} & 0.925611& -1.35025\times10^{-8} \\
 0 & 3.57819\times10^{-7} & 1.32895\times10^{-7} & 0.925613
\end{array}
\right),\non\\
&\bar{\eta}^{(2)}=\left(
\begin{array}{cccc}
 1 & 0 & 0 & 0 \\
 0 & 0.963414 & 0 & 0 \\
 0 & 0 & 0.963414 & 0 \\
 0 & 0 & 0 & 0.963414
\end{array}
\right),
\bar{\eta}^{(3)}=\left(
\begin{array}{cccc}
 1 & 0 & 0 & 0 \\
 0 & 0.990555 & 0 & 0 \\
 0 & 0 & 0.990555 & 0 \\
 0 & 0 & 0 & 0.990555
\end{array}
\right) .\non
\ee
Moreover, similar to Eq. (\ref{s3}), one can obtain,
\be
&\delta\eta^{(0)}=
\left(
\begin{array}{cccc}
 0 & 0 & 0 & 0 \\
 0 & 0.0298597 & 0.0258286 & 0.0258319 \\
 0 & 0.0259 & 0.0298154 & 0.0258115 \\
 0 & 0.0257602 & 0.025883 & 0.0297133
\end{array}
\right),
\delta\eta^{(1)}=\left(
\begin{array}{cccc}
 0 & 0 & 0 & 0 \\
 0 &0.0129621 & 0.0000229349 & 0.0000229077 \\
 0 & 0.0000229552 & 0.0129462 & 0.0000229223 \\
 0 & 0.0000226716 & 0.0000233086 & 0.0129497
\end{array}
\right),\non\\
&\delta\eta^{(2)}=\left(
\begin{array}{cccc}
 0 & 0 & 0 & 0 \\
 0 & 0.00533908 & 0 & 0 \\
 0 & 0 &0.00533904 & 0 \\
 0 & 0 & 0 &0.00533898
\end{array}
\right),
\delta\eta^{(3)}=\left(
\begin{array}{cccc}
 0 & 0 & 0 & 0 \\
 0 & 0.00125278 & 0 & 0 \\
 0 & 0 &0.00125278 & 0 \\
 0 & 0 & 0 &0.00125278
\end{array}
\right) .\non
\ee

(ii)  Consider the case where $F^{\mathrm{avg}(0)}=0.918$ ($f=0.94, k=0.05$), and the noise model of $\mathcal{N}$ is set as an amplitude damping noise. Similar to Eq. (\ref{s1}), one can obtain,
\be
&\eta^{\mathrm{avg}(0)}=
\left(
\begin{array}{cccc}
 1 & 0 & 0 & 0 \\
 0 & 0.908785 & 0 & 0 \\
 0 & 0 & 0.908785 & 0 \\
 0.112237 & 0 & 0 & 0.85443
\end{array}
\right),
\eta^{\mathrm{avg}(1)}=\left(
\begin{array}{cccc}
 1 & 0 & 0 & 0 \\
 0 & 0.92762 & 0 & 0 \\
 0 & 0 & 0.92762 & 0 \\
 -4.45261\times10^{-6} & 0 & 0 & 0.921814
\end{array}
\right),\non\\
&\eta^{\mathrm{avg}(2)}=\left(
\begin{array}{cccc}
 1 & 0 & 0 & 0 \\
 0 & 0.963508 & 0 & 0 \\
 0 & 0 & 0.963508 & 0 \\
 0 & 0 & 0 & 0.963436
\end{array}
\right),
\eta^{\mathrm{avg}(3)}=\left(
\begin{array}{cccc}
 1 & 0 & 0 & 0 \\
 0 & 0.990595 & 0 & 0 \\
 0 & 0 & 0.990595 & 0 \\
 0 & 0 & 0 & 0.990595
\end{array}
\right) .\non
\ee
Meanwhile, similar to Eq. (\ref{s2}), one can obtain,
\be
&\bar{\eta}^{(0)}=
\left(
\begin{array}{cccc}
 1 & 0 & 0 & 0 \\
 0 & 0.908843 & -0.000108809 & 0.000133326 \\
 0 & 0.000143222 & 0.90861 & -8.66705\times10^{-7} \\
 0.112237 & -0.0000533846 & -0.0000456332 & 0.854275
\end{array}
\right),\non\\
&\bar{\eta}^{(1)}=\left(
\begin{array}{cccc}
 1 & 0 & 0 & 0 \\
 -4.47766\times10^{-7} & 0.927539 & -2.2819\times10^{-7} & -1.09219\times10^{-6} \\
 4.21439\times10^{-7} & 2.30118\times10^{-9} & 0.927533 & -1.07714\times10^{-6} \\
 -3.66889\times10^{-6} & 1.27916\times10^{-6} & 1.09296\times10^{-6} & 0.9217
\end{array}
\right),\non\\
&\bar{\eta}^{(2)}=\left(
\begin{array}{cccc}
 1 & 0 & 0 & 0 \\
 0 & 0.963427 & 0 & 0 \\
 0 & 0 & 0.963427 & 0 \\
 0 & 0 & 0 & 0.963355
\end{array}
\right),
\bar{\eta}^{(3)}=\left(
\begin{array}{cccc}
 1 & 0 & 0 & 0 \\
 0 & 0.990554 & 0 & 0 \\
 0 & 0 & 0.990554 & 0 \\
 0 & 0 & 0 & 0.990554
\end{array}
\right) .\non
\ee
Moreover, similar to Eq. (\ref{s3}), one can obtain,
\be
&\delta\eta^{(0)}=
\left(
\begin{array}{cccc}
 0 & 0 & 0 & 0 \\
 0 & 0.0296858 & 0.025823 & 0.0259269 \\
 0 & 0.0257851 & 0.0299674 & 0.02579 \\
 0 & 0.0259645 & 0.0257521 & 0.0298341
\end{array}
\right),
\delta\eta^{(1)}=\left(
\begin{array}{cccc}
 0 & 0 & 0 & 0 \\
 0.0000699147 &0.0131219 & 0.0000230966 & 0.000151832 \\
 0.0000698661 & 0.0000229212 & 0.0131392 & 0.000152937 \\
 0.0000674748 & 0.000134124 & 0.000134663 & 0.0130151
\end{array}
\right),\non\\
&\delta\eta^{(2)}=\left(
\begin{array}{cccc}
 0 & 0 & 0 & 0 \\
 0 & 0.00525907 & 0 & 0 \\
 0 & 0 &0.00525921 & 0 \\
 0 & 0 & 0 &0.00525701
\end{array}
\right),
\delta\eta^{(3)}=\left(
\begin{array}{cccc}
 0 & 0 & 0 & 0 \\
 0 & 0.00118219 & 0 & 0 \\
 0 & 0 &0.00118219 & 0 \\
 0 & 0 & 0 &0.00118219
\end{array}
\right) .\non
\ee

(iii) Consider the case where $F^{\mathrm{avg}(0)}=0.918$ ($f=0.94, k=0.05$), and the noise model of $\mathcal{N}$ is set as an arbitrary numerical noise. The set of Kraus operators $\{A_{0},A_{1},A_{2},A_{3}\}$ are generated randomly as
\be
&A_{0}=\left(
\begin{array}{cc}
 0.55782 & 0.0184139 -0.0299884 i \\
 0.0184139 +0.0299884 i & 0.264979
\end{array}
\right),
A_{1}=\left(
\begin{array}{cc}
 0.0135882 -0.0221294 i & 0.0991524 +0.195463 i \\
 0.00307679 & -0.0135882+0.0221294 i
\end{array}
\right),\non\\
&A_{2}=\left(
\begin{array}{cc}
 0.00812856 +0.0132379 i & 0.00184055 \\
 0.0593135 -0.116927 i & -0.00812856-0.0132379 i
\end{array}
\right),
A_{3}=\left(
\begin{array}{cc}
 0.818093 & -0.00752438+0.012254 i \\
 -0.00752438-0.012254 i & 0.937755
\end{array}
\right).\non
\ee
Similar to Eq. (\ref{s1}), one can obtain,
\be
&\eta^{\mathrm{avg}(0)}=
\left(
\begin{array}{cccc}
 1 & 0 & 0 & 0 \\
 0.00368465 & 0.884802 & 0.000448593 & 0.00219029 \\
 0.00600072 & 0.000448593 & 0.885257 & 0.00356704 \\
 0.0292989 & 0.00219029 & 0.00356704 & 0.901943
\end{array}
\right),\non\\
&\eta^{\mathrm{avg}(1)}=\left(
\begin{array}{cccc}
 1 & 0 & 0 & 0 \\
 -4.52835\times10^{-6} & 0.92581 & 1.53159\times10^{-6} &2.1369\times10^{-6} \\
 -6.92152\times10^{-6} & 1.53082\times10^{-6} & 0.925841 & 3.28485\times10^{-6} \\
 -2.89255\times10^{-6} & 2.20405\times10^{-6} & 3.40018\times10^{-6} & 0.925252
\end{array}
\right),\non\\
&\eta^{\mathrm{avg}(2)}=\left(
\begin{array}{cccc}
 1 & 0 & 0 & 0 \\
 0 & 0.963438 & 0 & 0 \\
 0 & 0 & 0.963438 & 0 \\
 0 & 0 & 0 & 0.963438
\end{array}
\right),
\eta^{\mathrm{avg}(3)}=\left(
\begin{array}{cccc}
 1 & 0 & 0 & 0 \\
 0 & 0.990572 & 0 & 0 \\
 0 & 0 & 0.990572 & 0 \\
 0 & 0 & 0 & 0.990572
\end{array}
\right) .\non
\ee
Meanwhile, similar to Eq. (\ref{s2}), one can obtain,
\be
&\bar{\eta}^{(0)}=
\left(
\begin{array}{cccc}
 1 & 0 & 0 & 0 \\
 0.00368465 & 0.884772 & 0.00025835 & 0.00219715 \\
 0.00600072 & 0.000473319 & 0.885072 & 0.00366004 \\
 0.0292989 & 0.00200815 & 0.00360263 & 0.902019
\end{array}
\right),\non\\
&\bar{\eta}^{(1)}=\left(
\begin{array}{cccc}
 1 & 0 & 0 & 0 \\
 0.0000192865 & 0.925037 & 0.0000193851 & 0.0000198019 \\
 -0.0000194766 & 0.0000212745 & 0.925795 & -0.0000155139 \\
 -0.0000191295 &0.000019264 & -0.0000153212 & 0.926001
\end{array}
\right),\non\\
&\bar{\eta}^{(2)}=\left(
\begin{array}{cccc}
 1 & 0 & 0 & 0 \\
 0 & 0.963405 & 0 & 0 \\
 0 & 0 & 0.963407 & 0 \\
 0 & 0 & 0 & 0.963407
\end{array}
\right),
\bar{\eta}^{(3)}=\left(
\begin{array}{cccc}
 1 & 0 & 0 & 0 \\
 0 & 0.990558 & 0 & 0 \\
 0 & 0 & 0.990558 & 0 \\
 0 & 0 & 0 & 0.990558
\end{array}
\right) .\non
\ee
Moreover, similar to Eq. (\ref{s3}), one can obtain,
\be
&\delta\eta^{(0)}=
\left(
\begin{array}{cccc}
 0 & 0 & 0 & 0 \\
 0 & 0.0297525 & 0.0260041 & 0.0257256 \\
 0 & 0.0258935 & 0.0297769 & 0.0259085 \\
 0 & 0.0258368 & 0.0257975 & 0.0297762
\end{array}
\right),\non\\
&\delta\eta^{(1)}=\left(
\begin{array}{cccc}
 0 & 0 & 0 & 0 \\
 0.0000260314 &0.0128923 & 0.0000525842 & 0.0000522695 \\
 0.0000159412 & 0.0000525246 & 0.0128662 & 0.0000509399 \\
 0.0000189584 & 0.0000518149 & 0.0000504288 & 0.0129182
\end{array}
\right),\non\\
&\delta\eta^{(2)}=\left(
\begin{array}{cccc}
 0 & 0 & 0 & 0 \\
 0 &0.00538844 & 0 & 0 \\
 0 & 0 &0.00538858 & 0 \\
 0 & 0 & 0 &0.00538841
\end{array}
\right),
\delta\eta^{(3)}=\left(
\begin{array}{cccc}
 0 & 0 & 0 & 0 \\
 0 & 0.00119051 & 0 & 0 \\
 0 & 0 &0.00119051 & 0 \\
 0 & 0 & 0 &0.00119051
\end{array}
\right) .\non
\ee

\section{For 9-qubit code, some cases when average initial channel fidelity equals 0.9704}
\label{l6}

In our numerical simulations, it indicates that the effective channels are approximate to one of Pauli-$X$ and Pauli-$Z$ channels in each level, and in next concatenated level, the effective channels are approximate to the other. As an example, in the case where $f=0.98, k=0.02$, and $\mathcal{N}$ is fixed as amplitude damping noise, from the definition in Eq.~(\ref{b4}), the average QPMs in $l$-th level ($l=0,1,2,3$) are
\be
&\bar{\eta}^{(0)}=
\left(
\begin{array}{cccc}
 1 & 0 & 0 & 0 \\
 0 & 0.966927 & 0.0000251476&0.0000156073 \\
 0 & -4.08948\times10^{-6} & 0.966966 & -0.000038168 \\
 0.039002 & 8.68128\times10^{-6} & -0.0000162909 & 0.947644
\end{array}
\right),\non\\
&\bar{\eta}^{(1)}=\left(
\begin{array}{cccc}
  1 & 0 & 0 & 0 \\
 2.0898\times10^{-7} & 0.98151 & 0 & 0 \\
 0 & 0 & 0.976365 & 4.82745\times10^{-7} \\
 0 & 0 & -5.87662\times10^{-7} & 0.994331
\end{array}
\right),\non\\
&\bar{\eta}^{(2)}=\left(
\begin{array}{cccc}
 1 & 0 & 0 & 0 \\
 0 & 0.999856 & 0 & 0 \\
 0 & 0 & 0.995482 & 0 \\
 0 & 0 & 0 & 0.995624
\end{array}
\right),\bar{\eta}^{(3)}=\left(
\begin{array}{cccc}
 1 & 0 & 0 & 0 \\
 0 & 0.999914 & 0 & 0 \\
 0 & 0 & 0.999914 & 0 \\
 0 & 0 & 0 & 1
\end{array}
\right)  . \non
\ee
Moreover, from the definition in Eq.~(\ref{b5}), the SDs of elements of QPM in each concatenated level are
\be
&\delta\eta^{(0)}=
\left(
\begin{array}{cccc}
 0 & 0 & 0 & 0 \\
 0 & 0.0119553 & 0.0103716 & 0.010276 \\
 0 & 0.0103334 & 0.0118923 & 0.0103618 \\
 0 & 0.0103144 & 0.0103236 & 0.0119529
\end{array}
\right), \delta\eta^{(1)}=\left(
\begin{array}{cccc}
 0 & 0 & 0 & 0 \\
 4.58789\times10^{-8} &0.00175784 & 1.8762\times10^{-6} & 0 \\
 1.93633\times10^{-9} & 1.88581\times10^{-6} &0.00291518 & 0.0000551025 \\
 0 & 5.03582\times10^{-9} & 0.0000633488 & 0.00236694
\end{array}
\right),\non\\
&\delta\eta^{(2)}=\left(
\begin{array}{cccc}
 0 & 0 & 0 & 0 \\
 0 & 0.0000415446 & 0 & 0 \\
 0 & 0 &0.000281703 & 0 \\
 0 & 0 & 0 &0.000272073
\end{array}
\right), \delta\eta^{(3)}=\left(
\begin{array}{cccc}
 0 & 0 & 0 & 0 \\
 0 & 2.903\times10^{-6} & 0 & 0 \\
 0 & 0 &2.91276\times10^{-6} & 0 \\
 0 & 0 & 0 &5.71822\times10^{-8}
\end{array}
\right).\non
\ee

Similarly, for the case where $f=0.98, k=0.02$, and $\mathcal{N}$ is an arbitrarily generated numerical noise, the average QPMs in $l$-th level ($l=0,1,2,3$) are
\be
&\bar{\eta}^{(0)}=
\left(
\begin{array}{cccc}
 1 & 0 & 0 & 0 \\
 -0.0000108114 & 0.948612 & -0.0000450975&0.000694767 \\
 -3.72968\times10^{-6} & -6.47179\times10^{-6} & 0.948466 & 0.000374404 \\
 -0.000476769 & 0.000861642 & 0.000264038 & 0.984534
\end{array}
\right),\non\\
&\bar{\eta}^{(1)}=\left(
\begin{array}{cccc}
  1 & 0 & 0 & 0 \\
 0 & 0.998913 & 2.63852\times10^{-8} & 0 \\
 0 & -2.21341\times10^{-8} & 0.969132 & 1.33141\times10^{-9} \\
 0 & 0 &-1.39275\times10^{-9} & 0.970081
\end{array}
\right),\non\\
&\bar{\eta}^{(2)}=\left(
\begin{array}{cccc}
 1 & 0 & 0 & 0 \\
 0 & 0.996018 & 0 & 0 \\
 0 & 0 & 0.995986 & 0 \\
 0 & 0 & 0 & 0.999966
\end{array}
\right),\bar{\eta}^{(3)}=\left(
\begin{array}{cccc}
 1 & 0 & 0 & 0 \\
 0 & 1 & 0 & 0 \\
 0 & 0 & 0.999789 & 0 \\
 0 & 0 & 0 & 0.999789
\end{array}
\right)  . \non
\ee
The SDs of elements of QPM in each concatenated level are
\be
&\delta\eta^{(0)}=
\left(
\begin{array}{cccc}
 0 & 0 & 0 & 0 \\
 0 & 0.011925 & 0.0103388 & 0.0103094 \\
 0 & 0.0102404 & 0.0120343 & 0.0103647 \\
 0 & 0.0104072 & 0.0102665 & 0.0118446
\end{array}
\right), \delta\eta^{(1)}=\left(
\begin{array}{cccc}
 0 & 0 & 0 & 0 \\
 0 &0.000590257 & 1.92707\times10^{-6} & 0 \\
 0 & 1.71956\times10^{-6} &0.00425712 & 1.63097\times10^{-6} \\
 0 & 0 & 1.62991\times10^{-6} & 0.00404986
\end{array}
\right),\non\\
&\delta\eta^{(2)}=\left(
\begin{array}{cccc}
 0 & 0 & 0 & 0 \\
 0 & 0.000345456 & 0 & 0 \\
 0 & 0 &0.000349615 & 0 \\
 0 & 0 & 0 &9.42672\times10^{-6}
\end{array}
\right), \delta\eta^{(3)}=\left(
\begin{array}{cccc}
 0 & 0 & 0 & 0 \\
 0 & 1.02843\times10^{-9} & 0 & 0 \\
 0 & 0 &0.0000114166 & 0 \\
 0 & 0 & 0 &0.0000114161
\end{array}
\right).\non
\ee

\end{widetext}


\begin{references}
\bibitem{Shor} P. W. Shor, Phys. Rev. A {\bf52}, R2493(R) (1995).
\bibitem{Steane} A. M. Steane, Phys. Rev. Lett. {\bf77}, 793 (1996).
\bibitem{Bennett} C. H. Bennett, D. P. DiVincenzo, J. A. Smolin, and W. K.
Wootters, Phys. Rev. A {\bf54}, 3824 (1996).
\bibitem{Laflamme} R. Laflamme, C. Miquel, J. P. Paz, and W. H. Zurek, Phys. Rev. Lett. {\bf77}, 198 (1996).
\bibitem{KandL} E. Knill and R. Laflamme, Phys. Rev. A {\bf55}, 900 (1997).
\bibitem{Duan} L.-M. Duan and G.-C. Guo, Phys. Rev. Lett. {\bf79}, 1953 (1997).
\bibitem{Lidar} D. A. Lidar, I. L. Chuang, and K. B. Whaley, Phys. Rev. Lett. {\bf81}, 2594 (1998).
\bibitem{Zanardi} P. Zanardi and M. Rasetti, Phys. Rev. Lett. {\bf79}, 3306 (1997).
\bibitem{KandLV} E. Knill, R. Laflamme, and L. Viola, Phys. Rev. Lett. {\bf84}, 2525 (2000).
\bibitem{Zanardi2} P. Zanardi, Phys. Rev. A {\bf63}, 012301 (2000).
\bibitem{Kempe} J. Kempe, D. Bacon, D. A. Lidar, and K. B. Whaley, Phys. Rev. A {\bf63}, 042307 (2001).
\bibitem{Kribs} D. Kribs, R. Laflamme, and D. Poulin, Phys. Rev. Lett. {\bf94}, 180501 (2005).
\bibitem{Poulin 05} D. Poulin, Phys. Rev. Lett. {\bf95}, 230504 (2005).
\bibitem{Kribs2} D. W. Kribs and R. W. Spekkens, Phys. Rev. A {\bf74}, 042329 (2006).
\bibitem{Nielsen} M. A. Nielsen and I. L. Chuang, \emph{Quantum Computation and Quantum information}(Cambridge University Press,Cambridge, 2000).
\bibitem{Reimpell} M. Reimpell and R. F. Werner, Phys. Rev. Lett. {\bf94}, 080501 (2005).
\bibitem{Fletcher 07} A. S. Fletcher, P. W. Shor, and M. Z. Win, Phys. Rev. A {\bf75}, 012338 (2007).
\bibitem{Yamamoto} N. Yamamoto, S. Hara, and K. Tsumura, Phys. Rev. A {\bf71}, 022322 (2005).
\bibitem{Fletcher 08} A. S. Fletcher, P. W. Shor, and M. Z. Win, Phys. Rev. A {\bf77}, 012320 (2008).
\bibitem{Kosut PRL} R. L. Kosut, A. Shabani, and D. A. Lidar, Phys. Rev. Lett. {\bf100}, 020502 (2008).
\bibitem{Kosut} R. L. Kosut and D. A. Lidar, Quantum Inf. Proc. {\bf8}, 443 (2009).
\bibitem{Ball} G. Ball\'{o} and P. Gurin, Phys. Rev. A {\bf80}, 012326 (2009).
\bibitem{Poulin 06}  D. Poulin, Phys. Rev. A {\bf74}, 052333 (2006).
\bibitem{Rahn}  B. Rahn, A. C. Doherty, and H. Mabuchi, Phys. Rev. A {\bf66}, 032304 (2002).
\bibitem{Kesting} F. Kesting,  F. Fr\"{o}wis, and W. D\"{u}r, Phys. Rev. A {\bf88}, 042305 (2013).
\bibitem{PRL117.010501} C. Chamberland, T. Jochym-O'Connor, and R. Laflamme, Phys. Rev. Lett. {\bf117}, 010501 (2016).
\bibitem{PRA95.022313} C. Chamberland, T. Jochym-O'Connor, and R. Laflamme, Phys. Rev. A {\bf95}, 022313 (2017).
\bibitem{Theodore} T. J. Yoder, R. Takagi, and I. L. Chuang, Phys. Rev. X {\bf6}, 031039 (2016).
\bibitem{Ryuji} R. Takagi, T. J. Yoder, and I. L. Chuang, Phys. Rev. A {\bf96}, 042302 (2017).
\bibitem{Gilchrist} A. Gilchrist, N. K. Langford, and M. A. Nielsen, Phys. Rev. A {\bf71}, 062310 (2005).
\bibitem{Emerson} J. Emerson, M. Silva, O. Moussa, C. Ryan, M. Laforest, J. Baugh, D. G. Cory, and R. Laflamme, Science {\bf317}, 1893 (2007).
\bibitem{Knill 08} E. Knill, D. Leibfried, R. Reichle, J. Britton, R. B. Blakestad, J. D. Jost, C. Langer, R. Ozeri, S. Seidelin, and D. J. Wineland, Phys. Rev. A {\bf77}, 012307 (2008).
\bibitem{Bendersky} A. Bendersky, F. Pastawski, and J. P. Paz, Phys. Rev. A {\bf80}, 032116 (2009).
\bibitem{Magesan 11} E. Magesan, J. M. Gambetta, and J. Emerson, Phys. Rev. Lett. {\bf106}, 180504 (2011).
\bibitem{Magesan 12} E. Magesan, J. M. Gambetta, B. R. Johnson, C. A. Ryan, J. M. Chow, S. T. Merkel, M. P. da Silva, G. A. Keefe, M. B. Rothwell, and T. A. Ohki et al., Phys. Rev. Lett. {\bf109}, 080505 (2012).
\bibitem{HengYan} H.-Y. Wang, and W.-Q. Zheng. Sci. China-Phys. Mech. Astron. {\bf59}, 100313 (2016).
\bibitem{Ariano03}  G. M. D'Ariano, M. G. A.Paris, and M. F. Sacchi, Adv. Imaging. Electron. Phys. {\bf128}, 205 (2003).
\bibitem{Ariano04}  G. M. D'Ariano and J. LoPresti, in \emph{Quantum State Estimation}, edited by M. G. A. Paris and J. \v{R}eh\'{a}\v{c}ek. Lecture Notes in Physics, Vol. 649 (Springer, Berlin, 2004).
\bibitem{Chuang}  I. L. Chuang and M. A. Nielson, J. Mod. Opt. {\bf44}, 2455 (1997).
\bibitem{Poyatos}  J. F. Poyatos, J. I. Cirac, and P. Zoller, Phys. Rev. Lett. {\bf78}, 390 (1997).
\bibitem{Leung} D. W. Leung, Ph.D. thesis, Stanford University, 2000; G. M. D'Ariano and P. Lo Presti, Phys. Rev. Lett. {\bf86}, 4195 (2001); J. B. Altepeter, D. Branning, E. Jeffrey, T. C. Wei, P. G. Kwiat, R. T. Thew, J. L. O'Brien, M. A. Nielsen, and A. G. White, ibid. {\bf90}, 193601 (2003).
\bibitem{Childs} A. M. Childs, I. L. Chuang, and D. W. Leung, Phys. Rev. A {\bf64}, 012314 (2001).
\bibitem{Shao-Ming} S.-M. Fei, Sci. China-Phys. Mech. Astron. {\bf60}, 020331 (2017).
\bibitem{Huang} L. Huang, B. You, X. H. Wu, and T. Zhou, Phys. Rev. A {\bf92}, 052320 (2015).
\bibitem{C.C2017} C. Chamberland, J. Wallman, S. Beale, and R. Laflamme, Phys. Rev. A {\bf95}, 042332 (2017).
\bibitem{L. Steffen} L. Steffen, M. P. da Silva, A. Fedorov, M. Baur, and A. Wallraff, Phys. Rev. Lett. {\bf108}, 260506 (2012).
\bibitem{X.-H. Wu} X.-H. Wu and K. Xu, Quantum Inf. Proc. {\bf12}, 1379 (2013).
\bibitem{Schumacher} B. W. Schumacher, Phys. Rev. A {\bf54}, 2614 (1996).
\bibitem{Horodecki} M. Horodecki, P. Horodecki, and R. Horodecki, Phys. Rev. A {\bf60}, 1888 (1999).
\bibitem{Huang2}  L. Huang, X. H. Wu, and T. Zhou, arXiv: 1707.09788 (2017).
\bibitem{M.Guti} M. Guti\'{e}rrez, C. Smith, L. Lulushi, S. Janardan, and K. R. Brown, Phys. Rev. A {\bf94}, 042338 (2016).
\bibitem{S. J. Beale} S. J. Beale, J. J. Wallman, M. Guti\'{e}rrez, K. R. Brown, and R. Laflamme, Phys. Rev. Lett. {\bf121}, 190501 (2018).
\bibitem{D. Greenbaum} D. Greenbaum, Z. Dutton, Quantum Sci. Technol. {\bf3}, 015007 (2018).
\bibitem{E. Huang} E. Huang, A. C. Doherty, and S. Flammia, Phys. Rev. A {\bf99}, 022313 (2019).
\end{references}
\end{document}